\documentclass[english,prl,twocolumn,aps,superscriptaddress,longbibliography]{revtex4-2}
\usepackage[T1]{fontenc}
\usepackage[utf8]{inputenc}
\usepackage{color}
\usepackage{babel}
\usepackage{float}
\usepackage{amssymb}
\usepackage{graphicx}
\PassOptionsToPackage{normalem}{ulem}
\usepackage{ulem}
\usepackage[pdfusetitle,
 bookmarks=true,bookmarksnumbered=false,bookmarksopen=false,
 breaklinks=false,pdfborder={0 0 1},backref=false,colorlinks=true]
 {hyperref}
\hypersetup{
 linkcolor=Plum, urlcolor=violet, citecolor=Mulberry, pdfstartview={FitH}, hyperfootnotes=false, unicode=true}

\makeatletter

\@ifundefined{date}{}{\date{}}
\usepackage[dvipsnames]{xcolor}

\makeatother

\begin{document}
\preprint{APS/123-QED}
\title{Two fluctuating interfaces with sticking interactions:\\
Invariant measures and dynamics}
\author{Samvit Mahapatra}
\affiliation{Department of Physics, Ravenshaw University, Cuttack, Odisha 753003,
India}
\affiliation{School of Basic Sciences, Indian Institute of Technology Bhubaneswar,
Odisha 752050, India}
\author{Malay Bandyopadhyay}
\affiliation{School of Basic Sciences, Indian Institute of Technology Bhubaneswar,
Odisha 752050, India}
\author{Mustansir Barma}
\affiliation{Tata Institute of Fundamental Research, Gopanpally, Hyderabad 500046,
India}
\date{\today}
\begin{abstract}
We introduce and study a non-equilibrium stochastic model of two fluctuating
interfaces which interact through short-range attractive interactions
at their points of contact. Beginning from an entangled state, the
system exhibits diverse dynamics---ranging from fast transients with
small lifetimes to ultraslow evolution through quasi-stationary states---and
reaches stuck, entangled, or detached steady states. Near the stuck-detached
transition, two distinct dynamical modes of evolution co-occur. When
the two surfaces evolve through similar dynamics (both Edwards-Wilkinson
or both Kardar-Parisi-Zhang), the invariant measure is determined
and found to have an inhomogeneous product form. This exact steady
state is shown to be the measure of the equilibrium Poland-Scheraga
model of DNA denaturation.
\end{abstract}
\maketitle
\begin{figure*}
\includegraphics[scale=0.315]{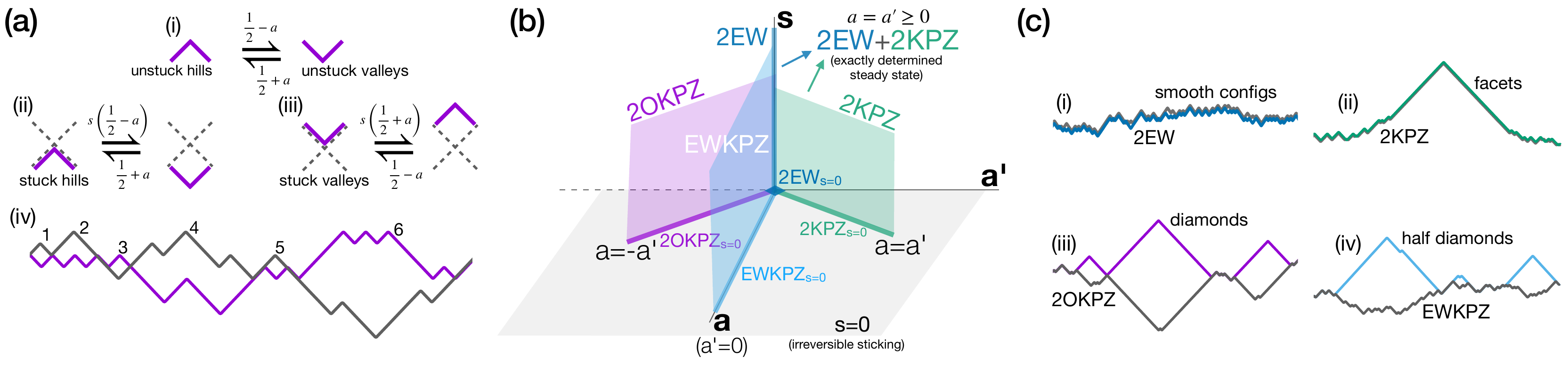}

\caption{(a) (i)-(iii) Update rules of one of the surfaces with bias $a$,
while the other surface has bias $a'$. The sticking interactions
are included by factoring the update probabilities of stuck sites
with a detachment probability $s$. (iv) A typical two-surface configuration
with its bubbles labelled as 1-6. (b) The four cases within the three-dimensional
phase diagram of the model in the space of parameters $a,\:a',\:s$.
(c) Schematic profiles of the transient structures that occur in the
different cases of the model. }\label{fig:phase=000020diagram}
\end{figure*}

\noindent Consider two thin 1-dimensional lines situated in close
proximity such that they interact with each other at their points
of contact or overlap, while fluctuating due to stochastic forces
or random environments. For example, we may imagine these fluctuating
lines as strings that are vibrating randomly, held together at their
points of contact, whereas segments away from contact fluctuate freely.
This general scenario connects to the subject of interacting random
walks \citep{Fisher1984WWWM}, and importantly to many different physical
systems, including interacting polymers \citep{RajasekaranSomendra1991DirectedWalksfixedpoint,Igloi1991TwoPolymers,NetzLipowsky1993threestrings,SomendraBaumgartner1997UnbindingPolymer,Yeomans1997DirectedPoly,Giacomin2007RandomPolymers},
double stranded DNA \citep{PolandScheraga1966Original,Fisher1966ExcludedVolume,KafriMukamelPeliti2000DNAfirstorder,CausoGrassberger2000DNAPerm,KafriMukamelPeliti2002DNAmelting,LubenskyNelson2000DNAunzipping,Marenzuddo2001DNAforce,MarenduzzoSomen2001DNAunzipping,CarlonStella2002StiffnessDNA,RichardGuttmann2004PolandScheraga,KafriPolkovnikov2006DNAUnzippingDPRM,Giacomin2007RandomPolymers,BarKafriMukamel2007Loop,BarKafriMukamel2008Melting,HankeMetzler2008StretchedDNA,Metzler2008DNABubble},
interacting vortex lines in superconductors \citep{Blatter1994RevModPhys,Tang1994TwoRepulsiveLines,halpin1995kinetic},
coupled wetting fronts \citep{Balankin2006PaperWetting}, interacting
ferromagnetic domain walls \citep{MetaxasPoliti2010CoupledDomainWalls,MetaxasPoliti2011CoupledDomainWalls},
and interacting lipid membranes \citep{Lipowsky1991Membranes,Lipowsky1996MembraneStickers}.
In the long time limit, such systems reach statistical steady states
with interesting properties, besides which there is great interest
in the manner in which these states are approached. Important questions
that arise are: Can steady state properties be calculated and characterized
analytically? How does the large-scale kinetics of approach depend
on the microscopic dynamics and initial conditions?

In this paper we answer these questions within a simple discrete stochastic
model of two randomly fluctuating strings with sticking interactions
at points of contact. We demonstrate the existence of both entangled
and detached steady states, both in and out of equilibrium, and transitions
between them as microscopic parameters are varied. Along specified
loci (including both equilibrium and nonequilibrium cases) we determine
the steady state exactly and show that it is described by the equilibrium
Poland-Scheraga measure. Further, for a wide class of parameters,
we show that there are two distinct modes of evolution, including
one which involves ultra-large time scales (growing super-exponentially
in the system size).

In our minimal model, each interface is taken to be a discrete single-step
surface in 1+1-dimensions \citep{SupplementalMaterial} with $L$
sites, described by heights $h_{ki}$, where the surfaces and their
sites are labelled by $k=1,2$ and $i=1,2,\:...\:L$, respectively.
Each surface satisfies the single-step constraint $|h_{ki}-h_{ki+1}|=1$
and periodic boundary conditions $h_{k1}=h_{kL}$. The surfaces are
initially entangled, and their configurations are allowed to cross
each other. The choice of single-step surfaces in 1+1-dimensions is
useful \citep{TangKrug1994PolymerConfinedExact,Krug1997AdvPhys} because
they are equivalent to the symmetric or asymmetric simple exclusion
process (SEP or ASEP) in 1-dimension; these models have been studied
extensively with a variety of analytical methods employed, yielding
rigorous results and exact solutions \citep{Spitzer1970ASEP,Derrida1998ReviewASEP,Derrida2007NonEqRev,BlytheEvans2007SolversGuide,SchadschneiderChowdhury2010StochasticTransport,Mallick2015ReviewASEP}.
Lowering the dimensionality of the problem simplifies its theoretical
analysis and computation, yet allows for systematic experimentation.
For example, the dynamics of stretched, isolated single molecules
of semi-flexible polymers such as double stranded DNA or actin has
been carefully studied under strong confinement \citep{Odijk1983ConfinedPolymers}
in experiments that include nano-channels (quasi 1-dimension) \citep{ReisnerFrey2005DNANanochannels,Indresh2020Nanochanneldynamics,Indresh2024Nanochanneldynamics}
and nano-slits (quasi 2-dimensions) \citep{Doyle2007DNASlitlike,TangDoyle2007Nanochannels,BalducciDoyle2008Nanoslit,Doyle2013IntrachainDNASlitlike},
or adsorbed on a 2-dimensional flat lipid membrane substrate \citep{MaierRadler1999DNALipids}.

The sticking interactions considered at the sites of contact embody
the effect of attractive interactions. They are incorporated through
the detachment probability $s$, a single parameter which satisfies
$0\le s\le1$. In a microscopic time step $[t,t+\Delta t]$, the heights
of the unstuck surface sites are updated as $h_{ki}\rightarrow h_{ki}\pm2$
with transition probabilities $(\frac{1}{2}\pm a)\Delta t$, while
the transition probabilities are $s(\frac{1}{2}\pm a)\Delta t$ at
sites where the surfaces cross each other (Fig. \ref{fig:phase=000020diagram}(a)).
Evidently, the interactions weaken as $s$ increases. Such contact
or short-range interactions between the interfaces constitute a special
case of $(h_{1},h_{2})$ couplings, with interaction determined by
the (difference of the) height fields of the interfaces. Equilibrium
models of coupled interfaces routinely carry short-range interactions;
the governing Hamiltonian incorporates them straightforwardly. The
analytical treatment of their non-equilibrium counterparts \citep{Barabasi1992CoupledInterfaces,Barabasi1993SurfactantCoupledInterfaces,ErtasKardar1992DirectedLines,ErtasKardar1993DriftingPolymers,SatyaDibyendu2005PersistenceCoupledInterfaces,Juntunen2007Interfaces,FerrariSpohn2013CoupledKPZ1D,MendlSpohn2013nlf-hydrodynamics,Spohn2014AnharmonicChains,SpohnStoltz2015coupledKPZ,SchutzKaufmann2017DdimDirectedPolymers,Bernardin2021coupledKPZBurgers,DeNardisGopalakrishnan2023SpinChainsKPZ,RoyDharSpohn2024coupledburgers,RoyDharSpohn2025coupledKPZ}
usually consider $(\nabla h_{1},\nabla h_{2})$ couplings, however,
in which case the interaction is determined by the local curvature
or height gradient fields of the two surfaces. The same also holds
for multicomponent lattice gases such as coupled ASEPs \citep{LR1997Sedimentation,LBR2000SPS,ArndtHeinzelRittenberg1998AHRoriginal,Rajewsky2000AHRRing,Popkov2014SuperdiffusiveModes,Popkov2015Fibonacci-universality,Chakraborty2017LH-statics,Chakraborty2017LH-dynamics,Mahapatra2020lightHeavy,Sasamoto2018Two-Species,Sasamoto2022Two-SpeciesAHR,SchmidtSchutzvanBeijeren2021ThreeLane,DolaiSimhaAbhikBasu2024CoupledASEP,CannizzaroOccelli2025FromABCtoKPZ,FerrariGernholt2025DecouplingDecayTasep,PrakashKabir2025scaledLH},
whose integrated density fields in 1-dimension correspond to the height
fields of random surfaces. Only a few non-equilibrium models have
been studied where $(h_{1},h_{2})$ couplings have been considered
\citep{Juntunen2007Interfaces,RoyDharSpohn2025coupledKPZ}, largely
numerically. 

We briefly highlight some interesting results, discussed in detail
later. 

-- The steady state is found exactly on a particular locus along
which the evolution rules of the two surfaces are identical -- either
linear diffusive Edwards-Wilkinson (EW) dynamics ($a=a'=0$), or nonlinear
Kardar-Parisi-Zhang (KPZ) dynamics ($a=a'\ne0$). The measure found
is the equilibrium Poland-Scheraga measure in both cases. 

-- Although the steady states are the same in the identical EW and
identical KPZ cases, the dynamics of approach to steady state is completely
different. If at least one of the surfaces evolves through KPZ dynamics
(and we have $a\ne a'$), two distinct modes of evolution occur depending
on the initial condition. One mode involves quick detachment of the
surfaces, whereas in the other, the surfaces get stuck tightly and
remain so until a rare fluctuation brings about their detachment.
We observe a regime where the growth of the steady state timescale
$T$ with system size $L$ is so large ($T\sim\exp(\lambda L^{2}$))
that it renders the steady state essentially unreachable. 

-- When both surfaces evolve through diffusive EW dynamics and stick
irreversibly, we numerically observe an atypical dynamic exponent
of $z\simeq1.5$. It also characterizes the scaling of the transients
when detachments are allowed.

With our dynamics the surfaces do not move together at the stuck segments.
For $s=0$, the two surfaces stick irreversibly, i.e. no detachments
are allowed. The steady state is a completely stuck, absorbing state
without any dynamics. In contrast, when detachments are allowed (non-zero
$s$), the steady state is a dynamic state. The two surfaces either
remain entangled with some degree of overlap, or detach completely.
The detached segments of the surfaces between two stuck segments are
generally termed as bubbles (shown in Fig. \ref{fig:phase=000020diagram}(a)).
For small, non-zero $s$, tiny bubbles can form and close, and across
the different cases we consider, the system shows interesting and
varied transients that reflect its dynamics for $s=0$. Evidently,
for $s=1$ the surfaces do not interact.

The biases of the two surfaces, $a$ and $a'$, and the detachment
probability $s$ together span the three-dimensional phase diagram
of the model (Fig. \ref{fig:phase=000020diagram}(b)). We present
results for each of four subspaces with distinctive features. 

(1) Two identical EW surfaces (2EW) ($a=a'=0$), 

(2) Two identical KPZ surfaces (2KPZ) ($a=a'\ne0$), 

(3) Two KPZ surfaces with drifts of equal magnitude 

but opposite orientation (2OKPZ) ($a=-a'\ne0$), 

(4) An EW surface and a KPZ surface (EWKPZ) 

($a=0,\;a'\ne0$). 

\uline{The steady state in the 2EW and 2KPZ cases}---Within the combined
2EW and 2KPZ subspace with $a=a'\ge0$ (Fig. 1b), we prove that pairwise
balance of probability fluxes is satisfied in the steady state \citep{Schutz1996Pairwise,TripathyBarma1997PRLDropPush,TripathyBarma1998PREDropPush};
its exact measure is determined for any non-zero $s$, and found to
be $a$-invariant. Recall that individual EW and KPZ surfaces possess
the same steady state measure despite strongly different dynamics
\citep{SupplementalMaterial}. We show that this property also extends
to our model, with two interacting surfaces.

\begin{figure}[h]
\includegraphics[scale=0.36]{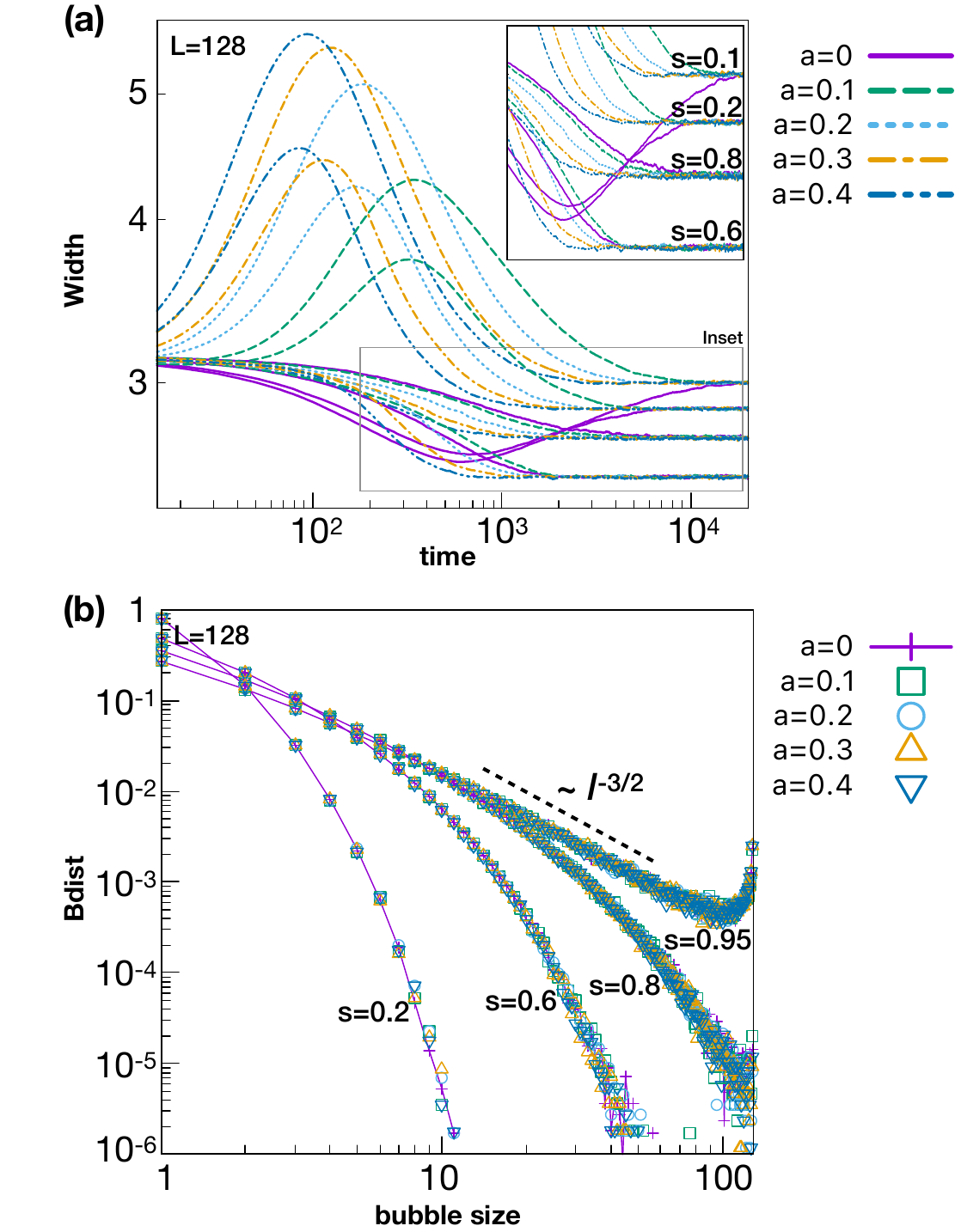}

\caption{(a) Invariance of the steady state width in the 2EW and 2KPZ cases
$a=a'$ for varying bias $a$. (b) The full size distributions of
the bubbles in steady state also reflect this invariance. The width
vs. time plots (bubble size distributions) are shown for detachment
probabilities $s=0.1,0.2,0.6,\mathrm{and}\:0.8$ ($s=0.2,0.6,0.8,\mathrm{and}\:0.95$).
For each value of $s$, the five plots of width (bubble size distributions)
with differing bias $a=0,0.1,0.2,0.3,\mathrm{and}\;0.4$ saturate
at a common width (superimpose on each other). The system size used
here is $L=128$ and the ensemble size is $N_{\mathrm{ens}}=50000$.
The variation of steady state width with detachment probability $s$
is non-monotonic for small-sized systems \citep{SupplementalMaterial}.
}\label{fig:2EW=0000202KPZ=000020Wss=000020invariance}
\end{figure}

Consider a configuration $C$ of the two surfaces. Let us denote the
outgoing (incoming) probability flux that exists from (to) configuration
$C$ to (from) another configuration $C'$ ($C''$) as $j_{C\rightarrow C'}$
($j_{C''\rightarrow C}$). In terms of incoming and outgoing transition
probability rates $w_{C''\rightarrow C}$ and $w_{C\rightarrow C'}$,
the fluxes are given by $j_{C\rightarrow C'}=w_{C\rightarrow C'}P(C)$,
and $j_{C''\rightarrow C}=w_{C''\rightarrow C}P(C'')$, respectively.
The master equation $d_{t}P(C)=\Sigma_{C''}j_{C''\rightarrow C}-\Sigma_{C'}j_{C\rightarrow C'}$
governs the time evolution of the probability $P(C)$ of configuration
$C$. In the 2EW and 2KPZ cases, when the individual incoming and
outgoing fluxes are paired in a specific way {[}see End Matter{]},
the master equation has the special form
\begin{eqnarray}
\frac{dP\left(C\right)}{dt} & = & \sum_{C',C''}A_{C}^{C',C''}\left(\overline{w}{}_{C''\rightarrow C}P\left(C''\right)-\overline{w}_{C\rightarrow C'}P\left(C\right)\right).\nonumber \\
\label{eq:MasterEqnSpl}
\end{eqnarray}
Here the transition probabilities $w$ have been decomposed by defining
$w=A\overline{w}$. The pairs of incoming-outgoing fluxes share the
same $a$-dependent coefficient $A\equiv A_{C}^{C',C''}\in\left\{ \frac{1}{2}+a,\frac{1}{2}-a\right\} $,
and $\overline{w}\in\left\{ 1,s\right\} $ takes the value $s$ for
fluxes involving a detachment, otherwise $\overline{w}=1$ (see Fig.
\ref{fig:pairwise=000020balance=000020illustration}). Upon imposing
that the paired fluxes balance each other $j_{C''\rightarrow C}=j_{C\rightarrow C'}$,
the steady state condition $\sum_{C''}j_{C''\rightarrow C}-\sum_{C'}j_{C\rightarrow C'}=0$
is ensured. Denoting a configuration with $k$ stuck sites by $C_{k}^{s}$,
the measure that supports the pairwise balance of fluxes has the form
\begin{eqnarray}
P\left(C_{k}^{s}\right) & = & p_{0}s^{L-k},\label{eq:P(Csk)}
\end{eqnarray}
for a specified value of $s<1$, where $p_{0}$ ($=1/4^{L}$) is the
measure of a completely stuck configuration. Therefore, the reduction
in transition probability by $s$ for every detachment is offset by
a corresponding increased probability of the configuration. The $a$-dependence
is confined only to the $A$'s and does not enter the probability
measure. This is validated strongly by numerical data of the saturation
width and stationary distribution of bubble sizes (Fig. \ref{fig:2EW=0000202KPZ=000020Wss=000020invariance}).
The probability measure is equiprobable within each set $\{C_{k}^{s}\}$
of all configurations with $k$ stuck sites, irrespective of the different
permutations and precise shapes of their bubbles and stuck segments.
In the thermodynamic limit the unnormalized probability measure for
any configuration $C$ has the product form $P(C)=p_{0}\prod_{i=1}^{L}p_{i}$,
with $p_{i}=1$ ($=s$) if the $i^{\mathrm{th}}$ site is stuck (detached).
The canonical partition function is given by\textcolor{blue}{{} }$Z(L)=4^{L}\sum_{k=1}^{L}P_{k}$,
where $P_{k}=\sum_{\{C_{k}^{s}\}}P(C_{k}^{s})=p_{0}s^{L-k}N(C_{k}^{s})$
is the marginal probability over the equiprobable set $\{C_{k}^{s}\}$
with $N(C_{k}^{s})$ configurations.

This steady state can be identified as the equilibrium measure of
the Poland-Scheraga model, a paradigmatic model of DNA denaturation
\citep{PolandScheraga1966Original,Fisher1966ExcludedVolume,KafriMukamelPeliti2000DNAfirstorder,RichardGuttmann2004PolandScheraga}.
The bubbles and stuck segments are uncorrelated in the measure described
by Eq. (\ref{eq:P(Csk)}). Accordingly, the probability of configuration
$C_{k}^{s}$ with $n$ bubbles (stuck segments) of length $l_{i}$
($r_{i}$) can be expressed as a product of the individual weights
of the alternating bubbles $B(l_{i})$ and stuck segments $R(r_{i})$,
along with sum constraints. This is given by $P(C_{k}^{s})\equiv P(l_{1},r_{1},l_{2},r_{2}\:...\:l_{n},r_{n})=4^{-L}\prod_{i=1}^{n}B(l_{i})R(r_{i})\,.\,\delta(L-\sum_{i}r_{i}+l_{i})\delta(k-\sum_{i}r_{i})$.
The weights $B(l)$ and $R(l)$ are determined from the number of
configurations of an individual bubble and a stuck segment, respectively.

Evidently, the stuck sites carry more weight ($p_{i}=1$) than the
detached sites that make up the bubbles ($p_{i}=s$). The bubbles,
however, have many more configurations than the stuck segments, as
in the PS model, attributing higher configurational entropy to the
bubbles. Counting the configurations of a bubble exactly, the leading
behavior of their weights is $B(l)\sim g^{l}/l^{c}$ in the limit
of large $l$, with $g=4s$, and the loop exponent $c=3/2$ \citep{SupplementalMaterial}.
The sum constraint yields $\Sigma_{i}^{n}l_{i}=L-k$, allowing the
factor $s^{L-k}$ in Eq. (\ref{eq:P(Csk)}) to be subsumed under the
weights $B(l)$. The weight of a stuck segment has the form $R(r)\sim v^{r}$
with $v=2$, because a stuck segment of length $r$ follows the trajectory
of a simple random walk, with $2^{r}$ different configurations. The
marginal probability becomes $P_{k}=4^{-L}v^{k}\sum\sum_{n,\{l_{i}\}}\prod_{i=1}^{n}B(l_{i})\:\delta(L-k-\sum_{i}l_{i})$\textcolor{blue}{.}
Subsequently, the partition function 
\begin{eqnarray}
Z(L) & = & \sum_{k}v^{k}g^{L-k}\sum_{n}\sum_{\{l_{i}\}}\prod_{i=1}^{n}\frac{1}{l_{i}^{c}}\:\delta(L-k-\sum_{i}l_{i})\label{eq:PartitionFn}
\end{eqnarray}
resembles that of the Poland-Scheraga model. 

In our model $g$ ($=4s$) is the control parameter with $0<s<1$,
and the value of $v$ is fixed. For $1\le c\le2$ the PS model shows
a continuous transition; the fraction of stuck DNA is non-zero for
$v>v_{c}$, with $v_{c}=g/1+\zeta(c)$, approaching zero continuously
in the limit $v\rightarrow v_{c}^{+}$. Our model corresponds to the
PS regime $v>v_{c}(s)$ for any $s$, indicating that the steady state
in the 2EW and 2KPZ subspaces is always entangled as $L\rightarrow\infty$.

\uline{Approach to steady state and transient dynamics}---We now
turn our attention to the second question at hand -- the approach
to the steady state and its dependence on the microscopic dynamics
and initial conditions. For each case, we present results from numerical
simulations and uncover important features and phenomena in the transient
dynamics. 

The initial state is drawn from an ensemble of random configurations
of the two surfaces, with equal heights at the first site, i.e. $h_{11}=h_{21}$,
ensuring some overlap in the beginning. A bubble in the initial state
is typically of size $\sim O(\sqrt{L})$. However, the largest bubble
in a configuration is of the order of the system size $\sim O(L)$,
reminiscent of the long leads in Gambler's Ruin \citep{SupplementalMaterial}. 

We monitor the width defined as $w(L,t)=\left\langle [\frac{1}{L}\sum_{i=1}^{n}(h_{i}(t)-\bar{h})^{2}]^{1/2}\right\rangle $,
with average instantaneous height $\bar{h}(t)=\frac{1}{L}\Sigma_{i=1}^{L}h_{i}(t)$.
In cases (1)-(3), the surfaces are statistically identical. It therefore
suffices to track the width of one of the surfaces. A second useful
observable is the fraction of surface sites that are stuck at any
given time -- the sticking fraction $\rho(t)=\frac{1}{L}\Sigma_{i=1}^{L}\delta\left(h_{1i}-h_{2i}\right)(t)$.
We also examine the size distribution of bubbles in some cases.

2EW and 2KPZ -- Smoothening and macroscopic facets: The closing of
the largest bubbles $\sim O(L)$ in the initial configurations dictates
the transient dynamics, while smaller bubbles $\sim O(\sqrt{L})$
close quickly.

In the 2EW case, the largest bubbles, while fluctuating diffusively,
close steadily from the edges like a zipper. The sticking process
progresses forward one site after another through diffusive fluctuations.
For $s=0$ the surfaces attain a relatively smooth final state compared
to the disordered initial state (see (i) in Fig. \ref{fig:phase=000020diagram}(c)). 

In the 2KPZ case, the surfaces undergo directed motion; the bubbles
close ballistically by drifting and deforming into sharp V-shaped
facets (see (ii) in Fig. \ref{fig:phase=000020diagram}(c)). Within
facets, the up and down tilts are cleanly phase separated. For $s=0$
the final state comprises an ensemble of configurations with macroscopic
facets arising from the largest bubbles. Transient facets are also
seen in models of wetting on attractive substrates \citep{Hinrichsen1997Wetting,Hinrichsen2000Wetting}.
\begin{figure*}
\includegraphics[scale=0.248]{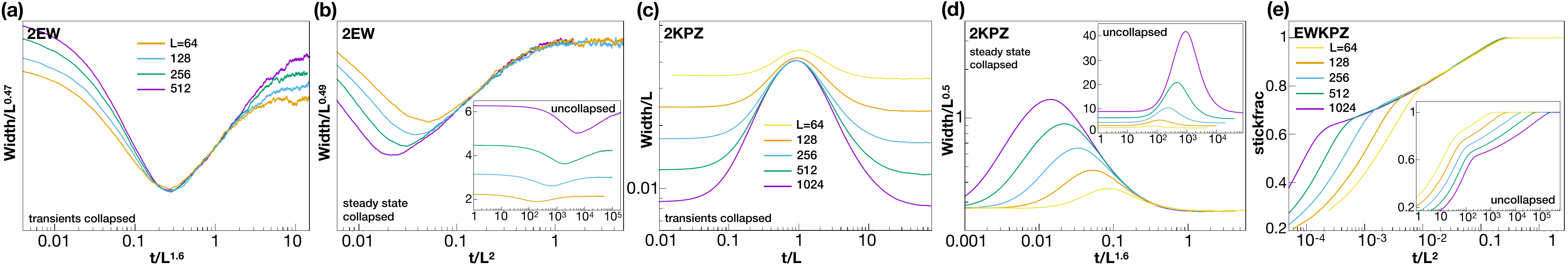}

\caption{(a) Collapse of the transient minimum in rescaled plots of width vs.
time in the 2EW case for different system sizes and small, non-zero
$s=0.1$. (b) Collapse of width at late times and steady state in
the 2EW case. (Inset) Uncollapsed plots. (c) Collapse of the transient
maximum in rescaled plots of width vs. time in the 2KPZ case for different
system sizes and $s=0.1$. (d) Collapse of width at late times and
steady state in the 2KPZ case. (Inset) Uncollapsed plots. (e) In the
EWKPZ case for no detachments allowed $s=0$, the sticking shows a
collapse when $S$ is plotted against $t/L^{z}$ with $z\simeq2$.
In (a)-(d) Log plots are shown for the collapsed data and semi-log
plots in the insets are shown for the uncollapsed data. Semi-log plots
are shown in (e). The plots are averaged over ensemble sizes $N_{\mathrm{ens}}\sim10^{4}$.
}\label{fig:=0000202EW=0000202KPZ=000020EWKPZ=000020scaling}
\end{figure*}

2EW and 2KPZ -- Scaling of transients and steady state: For small,
non-zero $s$, the two surfaces possess long-lived transient states
that are like the final states for $s=0$, aside from a few tiny bubbles.
Beyond these transients, the surfaces co-evolve further and undergo
roughening like a single fluctuating surface, forming and closing
tiny bubbles at different sites, and reach a dynamic steady state.

In the 2EW case, the surfaces attain a relatively smooth transient
state at intermediate times compared to their initial and final state;
accordingly the width shows a local minimum. The minimum width shows
a collapse when $W/L^{\alpha'}$ is plotted against $t/L^{\theta}$,
with $\alpha'\simeq0.46$ and $\theta\simeq1.6$ (Fig. \ref{fig:=0000202EW=0000202KPZ=000020EWKPZ=000020scaling}(a)),
where $L$ is the system size. The exponent $\theta$ corresponds
to the dynamic exponent $z$ for $s=0$, which is numerically found
to be $z\simeq1.5$ \citep{SupplementalMaterial}. The scaling behavior
at late times and steady state differs from the transients; the width
shows a collapse when $W/L^{\alpha}$ is plotted against $t/L^{z}$
with the same roughness exponent $\alpha\simeq0.5$ and dynamic exponent
$z\simeq2$ as a single EW surface (Fig. \ref{fig:=0000202EW=0000202KPZ=000020EWKPZ=000020scaling}(b)).

By contrast, the width of the surfaces in the 2KPZ case shows a local
maximum reflecting a faceted transient state. The maximum width shows
a collapse when the plots for different system sizes $L$ are rescaled
with exponents $\alpha'\simeq1$ and $\theta\simeq1$ (Fig. \ref{fig:=0000202EW=0000202KPZ=000020EWKPZ=000020scaling}(c)).
Beyond the maximum and in steady state, the width shows different
scaling as in the 2EW case, collapsing when rescaled using the same
exponents $\alpha\simeq0.5$ and $z\simeq1.6$ as a single KPZ surface
(Fig. \ref{fig:=0000202EW=0000202KPZ=000020EWKPZ=000020scaling}(d)). 

2OKPZ and EWKPZ -- Diamonds and half-diamonds: The bias(es) of the
KPZ surface(s) in the initial configurations is (are) oriented either
outward of the bubbles, or inward. The bubbles with bias(es) oriented
inward close quickly, while the bubbles with bias(es) oriented outward
quickly deform into faceted bubbles. 

\begin{figure}[h]
\includegraphics[scale=0.42]{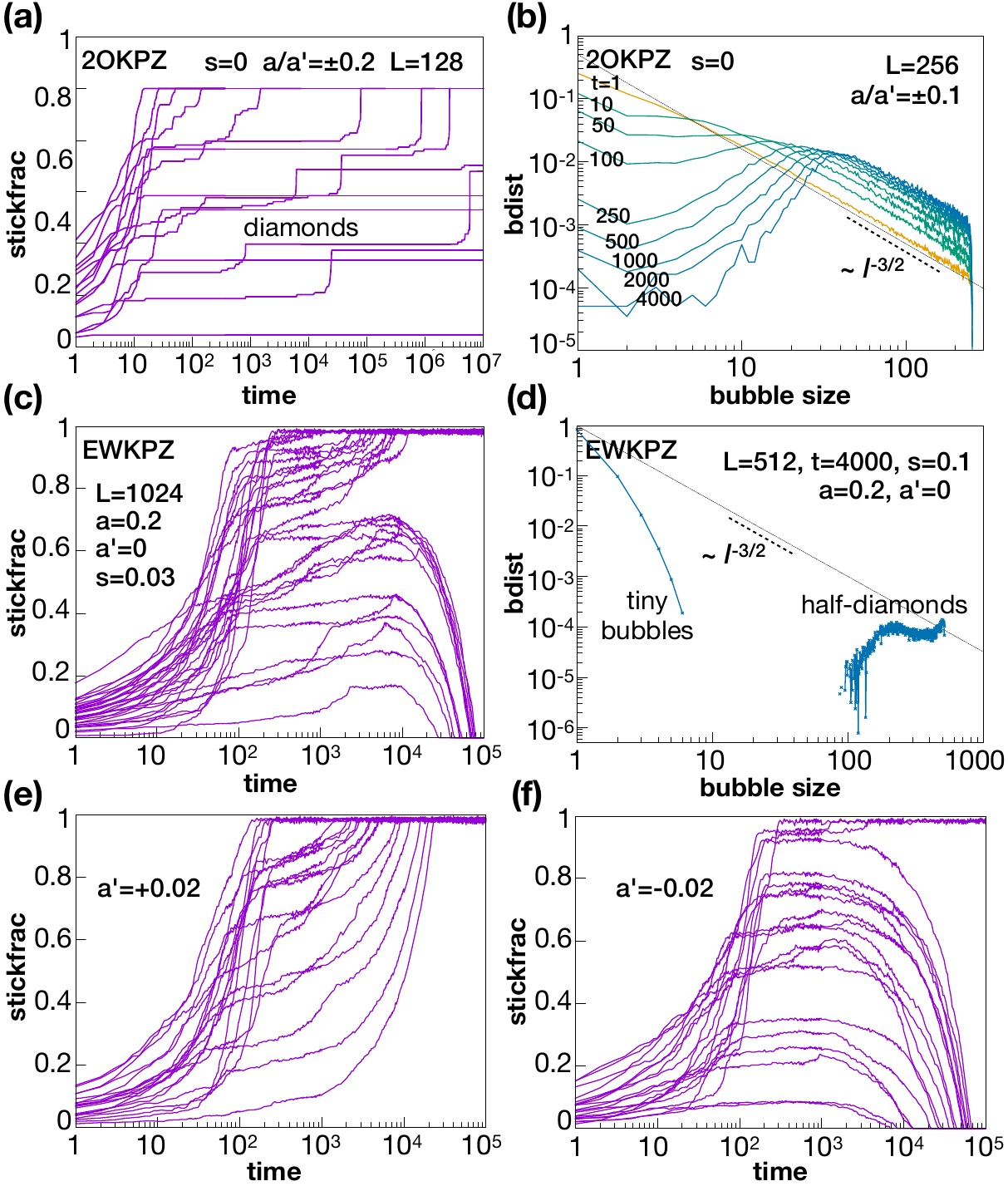}

\caption{In the 2OKPZ case for $s=0$, the sticking fraction vs. time trajectories
for individual configurations show (a) ultraslow evolution, because
diamonds persist without closing for several decades of time. (b)
With time the bubble size distribution recedes slowly from the small
$l$ region as the smaller diamonds close, while surviving diamonds
of size $l$ remain distributed as in the initial state $P(l)\sim l^{-3/2}$.
In the EWKPZ case for small non-zero $s$ ($=0.03$), the sticking
fraction trajectories show (c) a co-occurrence of two distinct modes
of evolution that lead to stuck and detached final configurations,
respectively, and (d) the size distributions of bubbles at relatively
large and finite times appear similar to real space condensates \citep{SatyaMajumdar2010Condensation}.
The distribution of smaller bubbles shows a decay, whereas the large,
growing half-diamonds in evolutions where the surfaces detach eventually
appear as a bump. The co-occurrence of evolutions vanishes when the
EW surface (with $a'=0$) in (c) is biased slightly; the surfaces
either attain (e) stuck configurations (shown for $a=+0.02$), or
(f) detach completely ($a'=-0.02$). }\label{fig:ultraslow=000020dynamics=000020and=000020coexistence}
\end{figure}

In the 2OKPZ case we observe diamond-shaped bubbles, hereafter known
as diamonds (see (iii) in Fig. \ref{fig:phase=000020diagram}(c)).
For $s=0$, the diamonds once formed are very stable and long-lived
with lifetimes of the largest bubbles being $\sim\exp\left(\lambda L^{2}\right)$
{[}see End Matter{]}. To put it succinctly, the diamonds are jammed
configurations due to the strong phase separation of their up and
down tilts. They possess almost no dynamics, except at the top-most
hill and lower-most valley sites, and close only through a slow deformation
process that proceeds against the biases of the two KPZ surfaces.
The ultraslow dynamics of diamonds is visible in the sticking fraction
vs time trajectories of individual configurations (Fig. \ref{fig:ultraslow=000020dynamics=000020and=000020coexistence}(a)),
and also in the slowly receding bubble size distribution (Fig. \ref{fig:ultraslow=000020dynamics=000020and=000020coexistence}(b)). 

Likewise in the EWKPZ case, the bubbles develop into half-diamonds
(see (iv) in Fig. \ref{fig:phase=000020diagram}(c)), as the KPZ surface
develops into a facet quickly and does not evolve any further, while
the EW surface continues fluctuating diffusively and subsequently
attaches onto the KPZ facet in timescales of $\sim O(L^{2})$. This
is visible in the plots of sticking fraction with time, which collapse
when plotted against $t/L^{z}$ with $z\simeq2$ (Fig. \ref{fig:=0000202EW=0000202KPZ=000020EWKPZ=000020scaling}(e)).
For $s=0$, the surfaces eventually reach a stuck faceted state, akin
to the 2KPZ case.

\begin{figure}[h]
\includegraphics[scale=0.35]{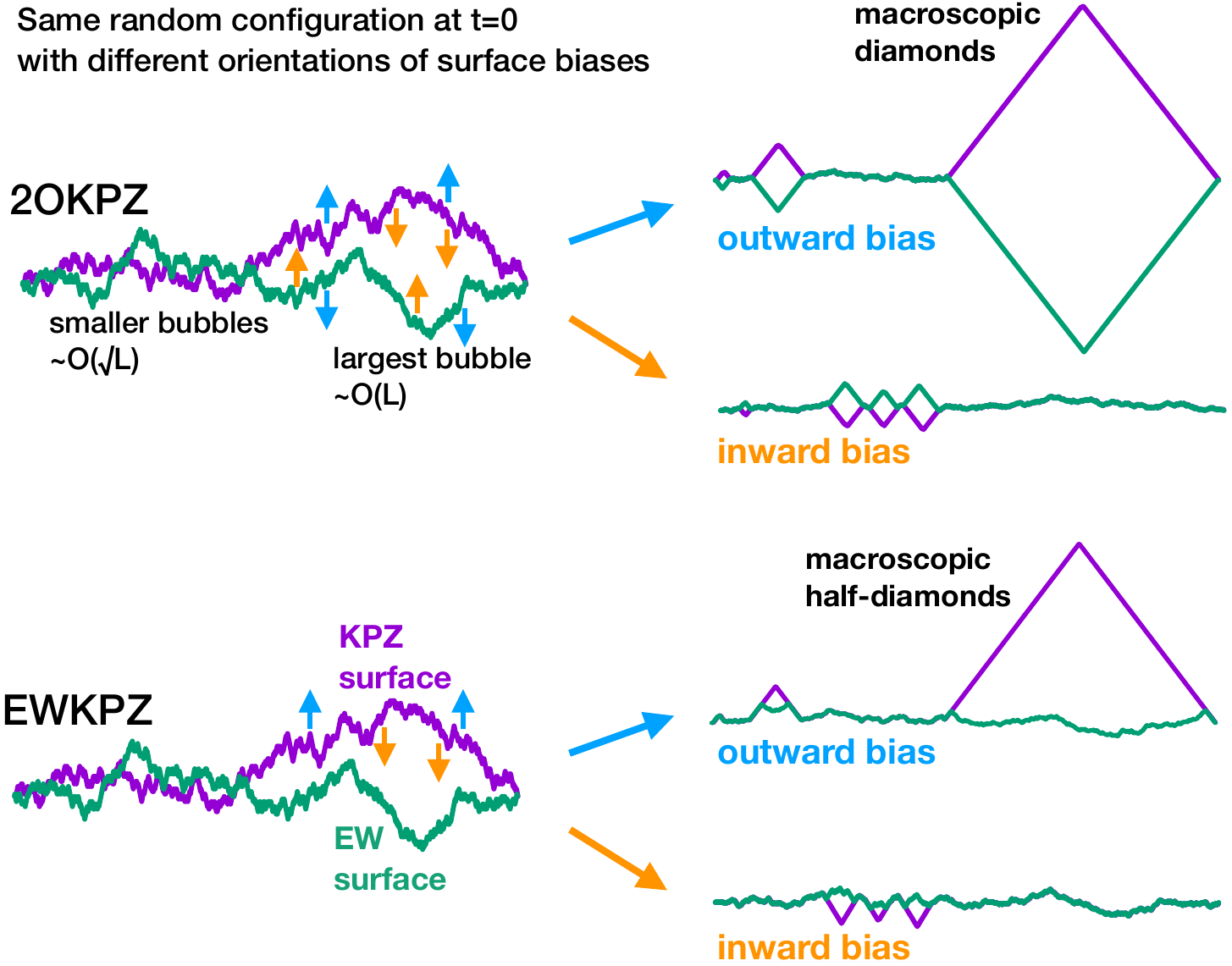}

\caption{In the EWKPZ case for small non-zero $s$ and the 2OKPZ case for small
$s$ and finite system size, we observe two distinct evolutions of
the two surfaces beginning from the same initial random configuration.
The differences in evolutions arise from the orientation of the bias(es)
of the KPZ surface(s) around the largest bubbles $\sim O(L)$ in the
initial configurations. The largest bubbles either collapse quickly,
or deform into half-diamonds and diamonds in the EWKPZ and 2OKPZ cases,
respectively. }\label{fig:schematic=000020two=000020distinct=000020evolutions}
\end{figure}

EWKPZ -- Two distinct evolutions: For small, non-zero $s$, the surfaces
in the EWKPZ case evolve in two distinct ways depending upon the orientation
of the KPZ bias around the largest bubbles $\sim O(L)$ in the initial
configuration (Fig. \ref{fig:schematic=000020two=000020distinct=000020evolutions}).
We observe (a) Evolutions where the KPZ surface is biased outwards;
the surfaces detach completely, passing through partially stuck transient
configurations with large half-diamonds. (b) Evolutions where the
KPZ surface is biased inwards; the largest bubbles close quickly,
and the surfaces get stuck near-completely and remain so indefinitely.
The two evolutions occur with probability of order $\sim O(1)$ within
the ensemble of evolving configurations \citep{SupplementalMaterial}.
Consequently, both nearly stuck and detached configurations of the
two surfaces occur in the steady state. This is observed in the sticking
fraction vs. time trajectories of individual configurations for large
$L$ and small $s$ (Fig. \ref{fig:ultraslow=000020dynamics=000020and=000020coexistence}(c)),
and also in the bubble size distributions at relatively large and
finite times (Fig. \ref{fig:ultraslow=000020dynamics=000020and=000020coexistence}(d)).
A heuristic explanation is presented in End Matter.

The EWKPZ case is the separatrix between the entangled and detached
states (Fig. \ref{fig:ultraslow=000020dynamics=000020and=000020coexistence}(e)-(f))
only at small non-zero $s$. When $s$ is relatively large the surfaces
predominantly detach in this case. The complete entangled-detached
separatrix for non-zero $s$ therefore remains to be evaluated. 

To summarize, we have introduced a minimal stochastic model of interacting
fluctuating interfaces with diverse forms of non-equilibrium behaviors
and steady state phases. The dynamics of the system allow understanding
and calculation of time-dependent properties. The model connects two
important classes of models in statistical physics, namely multicomponent
non-equilibrium systems and equilibrium models of DNA denaturation,
whose prototypical example is the Poland-Scheraga model. The origin
of the dynamic (transient) exponent $z\simeq1.5$ ($\theta\simeq1.6$)
observed in the 2EW case for $s=0$ \citep{SupplementalMaterial}
(non-zero $s$) remains to be understood. 

Acknowledgements---We acknowledge useful discussions with B.N. Sundaray,
D. Chaudhuri, I. Yadav, and S. Sen. S.M. would like to thank B.N.S.
for his valuable guidance during the research process. The simulations
were supported in part by SAMKHYA: HPC Facility provided by the Institute
of Physics, Bhubaneswar. S.M. was supported by the INSPIRE Fellowship
(IF190303) of the Department of Science and Technology, Government
of India. M.B. acknowledges the support of the Indian National Science
Academy (INSA), and the Department of Atomic Energy, Government of
India under Project Identification No. RTI 4007.

\bibliographystyle{apsrev4-2} 
\bibliography{TwoSurfaces_Outline_2024.bib}

\begin{thebibliography}{91}%
\makeatletter
\providecommand \@ifxundefined [1]{%
 \@ifx{#1\undefined}
}%
\providecommand \@ifnum [1]{%
 \ifnum #1\expandafter \@firstoftwo
 \else \expandafter \@secondoftwo
 \fi
}%
\providecommand \@ifx [1]{%
 \ifx #1\expandafter \@firstoftwo
 \else \expandafter \@secondoftwo
 \fi
}%
\providecommand \natexlab [1]{#1}%
\providecommand \enquote  [1]{``#1''}%
\providecommand \bibnamefont  [1]{#1}%
\providecommand \bibfnamefont [1]{#1}%
\providecommand \citenamefont [1]{#1}%
\providecommand \href@noop [0]{\@secondoftwo}%
\providecommand \href [0]{\begingroup \@sanitize@url \@href}%
\providecommand \@href[1]{\@@startlink{#1}\@@href}%
\providecommand \@@href[1]{\endgroup#1\@@endlink}%
\providecommand \@sanitize@url [0]{\catcode `\\12\catcode `\$12\catcode `\&12\catcode `\#12\catcode `\^12\catcode `\_12\catcode `\%12\relax}%
\providecommand \@@startlink[1]{}%
\providecommand \@@endlink[0]{}%
\providecommand \url  [0]{\begingroup\@sanitize@url \@url }%
\providecommand \@url [1]{\endgroup\@href {#1}{\urlprefix }}%
\providecommand \urlprefix  [0]{URL }%
\providecommand \Eprint [0]{\href }%
\providecommand \doibase [0]{https://doi.org/}%
\providecommand \selectlanguage [0]{\@gobble}%
\providecommand \bibinfo  [0]{\@secondoftwo}%
\providecommand \bibfield  [0]{\@secondoftwo}%
\providecommand \translation [1]{[#1]}%
\providecommand \BibitemOpen [0]{}%
\providecommand \bibitemStop [0]{}%
\providecommand \bibitemNoStop [0]{.\EOS\space}%
\providecommand \EOS [0]{\spacefactor3000\relax}%
\providecommand \BibitemShut  [1]{\csname bibitem#1\endcsname}%
\let\auto@bib@innerbib\@empty
\bibitem [{\citenamefont {Fisher}(1984)}]{Fisher1984WWWM}%
  \BibitemOpen
  \bibfield  {author} {\bibinfo {author} {\bibfnamefont {M.~E.}\ \bibnamefont {Fisher}},\ }\href {https://doi.org/https://doi.org/10.1007/BF01009436} {\bibfield  {journal} {\bibinfo  {journal} {J. Stat. Phys}\ }\textbf {\bibinfo {volume} {34}},\ \bibinfo {pages} {667} (\bibinfo {year} {1984})}\BibitemShut {NoStop}%
\bibitem [{\citenamefont {Rajasekaran}\ and\ \citenamefont {Bhattacharjee}(1991)}]{RajasekaranSomendra1991DirectedWalksfixedpoint}%
  \BibitemOpen
  \bibfield  {author} {\bibinfo {author} {\bibfnamefont {J.}~\bibnamefont {Rajasekaran}}\ and\ \bibinfo {author} {\bibfnamefont {S.~M.}\ \bibnamefont {Bhattacharjee}},\ }\href {https://doi.org/10.1088/0305-4470/24/7/010} {\bibfield  {journal} {\bibinfo  {journal} {J. Phys. A: Math. Gen.}\ }\textbf {\bibinfo {volume} {24}},\ \bibinfo {pages} {L371} (\bibinfo {year} {1991})}\BibitemShut {NoStop}%
\bibitem [{\citenamefont {Igloi}(1991)}]{Igloi1991TwoPolymers}%
  \BibitemOpen
  \bibfield  {author} {\bibinfo {author} {\bibfnamefont {F.}~\bibnamefont {Igloi}},\ }\href {https://doi.org/10.1209/0295-5075/16/2/009} {\bibfield  {journal} {\bibinfo  {journal} {Europhys. Lett.}\ }\textbf {\bibinfo {volume} {16}},\ \bibinfo {pages} {171} (\bibinfo {year} {1991})}\BibitemShut {NoStop}%
\bibitem [{\citenamefont {Netz}\ and\ \citenamefont {Lipowsky}(1993)}]{NetzLipowsky1993threestrings}%
  \BibitemOpen
  \bibfield  {author} {\bibinfo {author} {\bibfnamefont {R.~R.}\ \bibnamefont {Netz}}\ and\ \bibinfo {author} {\bibfnamefont {R.}~\bibnamefont {Lipowsky}},\ }\href {https://doi.org/10.1103/PhysRevE.47.3039} {\bibfield  {journal} {\bibinfo  {journal} {Phys. Rev. E}\ }\textbf {\bibinfo {volume} {47}},\ \bibinfo {pages} {3039} (\bibinfo {year} {1993})}\BibitemShut {NoStop}%
\bibitem [{\citenamefont {Bhattacharjee}\ and\ \citenamefont {Baumg{\"a}rtner}(1997)}]{SomendraBaumgartner1997UnbindingPolymer}%
  \BibitemOpen
  \bibfield  {author} {\bibinfo {author} {\bibfnamefont {S.~M.}\ \bibnamefont {Bhattacharjee}}\ and\ \bibinfo {author} {\bibfnamefont {A.}~\bibnamefont {Baumg{\"a}rtner}},\ }\href {https://doi.org/10.1063/1.474995} {\bibfield  {journal} {\bibinfo  {journal} {J. Chem. Phys.}\ }\textbf {\bibinfo {volume} {107}},\ \bibinfo {pages} {7571} (\bibinfo {year} {1997})}\BibitemShut {NoStop}%
\bibitem [{\citenamefont {Yeomans}(1997)}]{Yeomans1997DirectedPoly}%
  \BibitemOpen
  \bibfield  {author} {\bibinfo {author} {\bibfnamefont {J.~M.}\ \bibnamefont {Yeomans}},\ }\bibinfo {title} {Directed-walk models of polymers and wetting},\ in\ \href {https://doi.org/10.1017/CBO9780511564284.022} {\emph {\bibinfo {booktitle} {Nonequilibrium Statistical Mechanics in One Dimension}}},\ \bibinfo {editor} {edited by\ \bibinfo {editor} {\bibfnamefont {V.}~\bibnamefont {Privman}}}\ (\bibinfo  {publisher} {Cambridge University Press},\ \bibinfo {year} {1997})\ p.\ \bibinfo {pages} {329–334}\BibitemShut {NoStop}%
\bibitem [{\citenamefont {Giacomin}(2007)}]{Giacomin2007RandomPolymers}%
  \BibitemOpen
  \bibfield  {author} {\bibinfo {author} {\bibfnamefont {G.}~\bibnamefont {Giacomin}},\ }\href {https://doi.org/10.1142/p504} {\emph {\bibinfo {title} {Random Polymer Models}}}\ (\bibinfo  {publisher} {Imperial College Press},\ \bibinfo {year} {2007})\BibitemShut {NoStop}%
\bibitem [{\citenamefont {Poland}\ and\ \citenamefont {Scheraga}(1966)}]{PolandScheraga1966Original}%
  \BibitemOpen
  \bibfield  {author} {\bibinfo {author} {\bibfnamefont {D.}~\bibnamefont {Poland}}\ and\ \bibinfo {author} {\bibfnamefont {H.~A.}\ \bibnamefont {Scheraga}},\ }\href {https://doi.org/https://doi.org/10.1063/1.1727786} {\bibfield  {journal} {\bibinfo  {journal} {J. Chem. Phys.}\ }\textbf {\bibinfo {volume} {45}},\ \bibinfo {pages} {1464} (\bibinfo {year} {1966})}\BibitemShut {NoStop}%
\bibitem [{\citenamefont {Fisher}(1966)}]{Fisher1966ExcludedVolume}%
  \BibitemOpen
  \bibfield  {author} {\bibinfo {author} {\bibfnamefont {M.~E.}\ \bibnamefont {Fisher}},\ }\href {https://doi.org/https://doi.org/10.1063/1.1727787} {\bibfield  {journal} {\bibinfo  {journal} {J. Chem. Phys.}\ }\textbf {\bibinfo {volume} {45}},\ \bibinfo {pages} {1469} (\bibinfo {year} {1966})}\BibitemShut {NoStop}%
\bibitem [{\citenamefont {Kafri}\ \emph {et~al.}(2000)\citenamefont {Kafri}, \citenamefont {Mukamel},\ and\ \citenamefont {Peliti}}]{KafriMukamelPeliti2000DNAfirstorder}%
  \BibitemOpen
  \bibfield  {author} {\bibinfo {author} {\bibfnamefont {Y.}~\bibnamefont {Kafri}}, \bibinfo {author} {\bibfnamefont {D.}~\bibnamefont {Mukamel}},\ and\ \bibinfo {author} {\bibfnamefont {L.}~\bibnamefont {Peliti}},\ }\href {https://doi.org/10.1103/PhysRevLett.85.4988} {\bibfield  {journal} {\bibinfo  {journal} {Phys. Rev. Lett.}\ }\textbf {\bibinfo {volume} {85}},\ \bibinfo {pages} {4988} (\bibinfo {year} {2000})}\BibitemShut {NoStop}%
\bibitem [{\citenamefont {Causo}\ \emph {et~al.}(2000)\citenamefont {Causo}, \citenamefont {Coluzzi},\ and\ \citenamefont {Grassberger}}]{CausoGrassberger2000DNAPerm}%
  \BibitemOpen
  \bibfield  {author} {\bibinfo {author} {\bibfnamefont {M.~S.}\ \bibnamefont {Causo}}, \bibinfo {author} {\bibfnamefont {B.}~\bibnamefont {Coluzzi}},\ and\ \bibinfo {author} {\bibfnamefont {P.}~\bibnamefont {Grassberger}},\ }\href {https://doi.org/10.1103/PhysRevE.62.3958} {\bibfield  {journal} {\bibinfo  {journal} {Phys. Rev. E}\ }\textbf {\bibinfo {volume} {62}},\ \bibinfo {pages} {3958} (\bibinfo {year} {2000})}\BibitemShut {NoStop}%
\bibitem [{\citenamefont {Kafri}\ \emph {et~al.}(2002)\citenamefont {Kafri}, \citenamefont {Mukamel},\ and\ \citenamefont {Peliti}}]{KafriMukamelPeliti2002DNAmelting}%
  \BibitemOpen
  \bibfield  {author} {\bibinfo {author} {\bibfnamefont {Y.}~\bibnamefont {Kafri}}, \bibinfo {author} {\bibfnamefont {D.}~\bibnamefont {Mukamel}},\ and\ \bibinfo {author} {\bibfnamefont {L.}~\bibnamefont {Peliti}},\ }\href {https://doi.org/10.1140/epjb/e20020138} {\bibfield  {journal} {\bibinfo  {journal} {Eur. Phys. J. B}\ }\textbf {\bibinfo {volume} {27}},\ \bibinfo {pages} {135} (\bibinfo {year} {2002})}\BibitemShut {NoStop}%
\bibitem [{\citenamefont {Lubensky}\ and\ \citenamefont {Nelson}(2000)}]{LubenskyNelson2000DNAunzipping}%
  \BibitemOpen
  \bibfield  {author} {\bibinfo {author} {\bibfnamefont {D.~K.}\ \bibnamefont {Lubensky}}\ and\ \bibinfo {author} {\bibfnamefont {D.~R.}\ \bibnamefont {Nelson}},\ }\href {https://doi.org/10.1103/PhysRevLett.85.1572} {\bibfield  {journal} {\bibinfo  {journal} {Phys. Rev. Lett.}\ }\textbf {\bibinfo {volume} {85}},\ \bibinfo {pages} {1572} (\bibinfo {year} {2000})}\BibitemShut {NoStop}%
\bibitem [{\citenamefont {Marenduzzo}\ \emph {et~al.}(2001{\natexlab{a}})\citenamefont {Marenduzzo}, \citenamefont {Trovato},\ and\ \citenamefont {Maritan}}]{Marenzuddo2001DNAforce}%
  \BibitemOpen
  \bibfield  {author} {\bibinfo {author} {\bibfnamefont {D.}~\bibnamefont {Marenduzzo}}, \bibinfo {author} {\bibfnamefont {A.}~\bibnamefont {Trovato}},\ and\ \bibinfo {author} {\bibfnamefont {A.}~\bibnamefont {Maritan}},\ }\href {https://doi.org/10.1103/PhysRevE.64.031901} {\bibfield  {journal} {\bibinfo  {journal} {Phys. Rev. E}\ }\textbf {\bibinfo {volume} {64}},\ \bibinfo {pages} {031901} (\bibinfo {year} {2001}{\natexlab{a}})}\BibitemShut {NoStop}%
\bibitem [{\citenamefont {Marenduzzo}\ \emph {et~al.}(2001{\natexlab{b}})\citenamefont {Marenduzzo}, \citenamefont {Bhattacharjee}, \citenamefont {Maritan}, \citenamefont {Orlandini},\ and\ \citenamefont {Seno}}]{MarenduzzoSomen2001DNAunzipping}%
  \BibitemOpen
  \bibfield  {author} {\bibinfo {author} {\bibfnamefont {D.}~\bibnamefont {Marenduzzo}}, \bibinfo {author} {\bibfnamefont {S.~M.}\ \bibnamefont {Bhattacharjee}}, \bibinfo {author} {\bibfnamefont {A.}~\bibnamefont {Maritan}}, \bibinfo {author} {\bibfnamefont {E.}~\bibnamefont {Orlandini}},\ and\ \bibinfo {author} {\bibfnamefont {F.}~\bibnamefont {Seno}},\ }\href {https://doi.org/10.1103/PhysRevLett.88.028102} {\bibfield  {journal} {\bibinfo  {journal} {Phys. Rev. Lett.}\ }\textbf {\bibinfo {volume} {88}},\ \bibinfo {pages} {028102} (\bibinfo {year} {2001}{\natexlab{b}})}\BibitemShut {NoStop}%
\bibitem [{\citenamefont {Carlon}\ \emph {et~al.}(2002)\citenamefont {Carlon}, \citenamefont {Orlandini},\ and\ \citenamefont {Stella}}]{CarlonStella2002StiffnessDNA}%
  \BibitemOpen
  \bibfield  {author} {\bibinfo {author} {\bibfnamefont {E.}~\bibnamefont {Carlon}}, \bibinfo {author} {\bibfnamefont {E.}~\bibnamefont {Orlandini}},\ and\ \bibinfo {author} {\bibfnamefont {A.~L.}\ \bibnamefont {Stella}},\ }\href {https://doi.org/10.1103/PhysRevLett.88.198101} {\bibfield  {journal} {\bibinfo  {journal} {Phys. Rev. Lett.}\ }\textbf {\bibinfo {volume} {88}},\ \bibinfo {pages} {198101} (\bibinfo {year} {2002})}\BibitemShut {NoStop}%
\bibitem [{\citenamefont {Richard}\ and\ \citenamefont {Guttmann}(2004)}]{RichardGuttmann2004PolandScheraga}%
  \BibitemOpen
  \bibfield  {author} {\bibinfo {author} {\bibfnamefont {C.}~\bibnamefont {Richard}}\ and\ \bibinfo {author} {\bibfnamefont {A.~J.}\ \bibnamefont {Guttmann}},\ }\href {https://doi.org/https://doi.org/10.1023/B:JOSS.0000022370.48118.8b} {\bibfield  {journal} {\bibinfo  {journal} {J. Stat. Phys.}\ }\textbf {\bibinfo {volume} {115}},\ \bibinfo {pages} {925} (\bibinfo {year} {2004})}\BibitemShut {NoStop}%
\bibitem [{\citenamefont {Kafri}\ and\ \citenamefont {Polkovnikov}(2006)}]{KafriPolkovnikov2006DNAUnzippingDPRM}%
  \BibitemOpen
  \bibfield  {author} {\bibinfo {author} {\bibfnamefont {Y.}~\bibnamefont {Kafri}}\ and\ \bibinfo {author} {\bibfnamefont {A.}~\bibnamefont {Polkovnikov}},\ }\href {https://doi.org/10.1103/PhysRevLett.97.208104} {\bibfield  {journal} {\bibinfo  {journal} {Phys. Rev. Lett.}\ }\textbf {\bibinfo {volume} {97}},\ \bibinfo {pages} {208104} (\bibinfo {year} {2006})}\BibitemShut {NoStop}%
\bibitem [{\citenamefont {Bar}\ \emph {et~al.}(2007)\citenamefont {Bar}, \citenamefont {Kafri},\ and\ \citenamefont {Mukamel}}]{BarKafriMukamel2007Loop}%
  \BibitemOpen
  \bibfield  {author} {\bibinfo {author} {\bibfnamefont {A.}~\bibnamefont {Bar}}, \bibinfo {author} {\bibfnamefont {Y.}~\bibnamefont {Kafri}},\ and\ \bibinfo {author} {\bibfnamefont {D.}~\bibnamefont {Mukamel}},\ }\href {https://doi.org/10.1103/PhysRevLett.98.038103} {\bibfield  {journal} {\bibinfo  {journal} {Phys. Rev. Lett.}\ }\textbf {\bibinfo {volume} {98}},\ \bibinfo {pages} {038103} (\bibinfo {year} {2007})}\BibitemShut {NoStop}%
\bibitem [{\citenamefont {Bar}\ \emph {et~al.}(2008)\citenamefont {Bar}, \citenamefont {Kafri},\ and\ \citenamefont {Mukamel}}]{BarKafriMukamel2008Melting}%
  \BibitemOpen
  \bibfield  {author} {\bibinfo {author} {\bibfnamefont {A.}~\bibnamefont {Bar}}, \bibinfo {author} {\bibfnamefont {Y.}~\bibnamefont {Kafri}},\ and\ \bibinfo {author} {\bibfnamefont {D.}~\bibnamefont {Mukamel}},\ }\href {https://doi.org/10.1088/0953-8984/21/3/034110} {\bibfield  {journal} {\bibinfo  {journal} {J. Phys.: Condens. Matter}\ }\textbf {\bibinfo {volume} {21}},\ \bibinfo {pages} {034110} (\bibinfo {year} {2008})}\BibitemShut {NoStop}%
\bibitem [{\citenamefont {Hanke}\ \emph {et~al.}(2008)\citenamefont {Hanke}, \citenamefont {Ochoa},\ and\ \citenamefont {Metzler}}]{HankeMetzler2008StretchedDNA}%
  \BibitemOpen
  \bibfield  {author} {\bibinfo {author} {\bibfnamefont {A.}~\bibnamefont {Hanke}}, \bibinfo {author} {\bibfnamefont {M.~G.}\ \bibnamefont {Ochoa}},\ and\ \bibinfo {author} {\bibfnamefont {R.}~\bibnamefont {Metzler}},\ }\href {https://doi.org/10.1103/PhysRevLett.100.018106} {\bibfield  {journal} {\bibinfo  {journal} {Phys. Rev. Lett.}\ }\textbf {\bibinfo {volume} {100}},\ \bibinfo {pages} {018106} (\bibinfo {year} {2008})}\BibitemShut {NoStop}%
\bibitem [{\citenamefont {Metzler}\ \emph {et~al.}(2008)\citenamefont {Metzler}, \citenamefont {Ambj{\"o}rnsson}, \citenamefont {Hanke},\ and\ \citenamefont {Fogedby}}]{Metzler2008DNABubble}%
  \BibitemOpen
  \bibfield  {author} {\bibinfo {author} {\bibfnamefont {R.}~\bibnamefont {Metzler}}, \bibinfo {author} {\bibfnamefont {T.}~\bibnamefont {Ambj{\"o}rnsson}}, \bibinfo {author} {\bibfnamefont {A.}~\bibnamefont {Hanke}},\ and\ \bibinfo {author} {\bibfnamefont {H.~C.}\ \bibnamefont {Fogedby}},\ }\href {https://doi.org/10.1088/0953-8984/21/3/034111} {\bibfield  {journal} {\bibinfo  {journal} {J. Phys. Condens. Matter}\ }\textbf {\bibinfo {volume} {21}},\ \bibinfo {pages} {034111} (\bibinfo {year} {2008})}\BibitemShut {NoStop}%
\bibitem [{\citenamefont {Blatter}\ \emph {et~al.}(1994)\citenamefont {Blatter}, \citenamefont {Feigel'man}, \citenamefont {Geshkenbein}, \citenamefont {Larkin},\ and\ \citenamefont {Vinokur}}]{Blatter1994RevModPhys}%
  \BibitemOpen
  \bibfield  {author} {\bibinfo {author} {\bibfnamefont {G.}~\bibnamefont {Blatter}}, \bibinfo {author} {\bibfnamefont {M.~V.}\ \bibnamefont {Feigel'man}}, \bibinfo {author} {\bibfnamefont {V.~B.}\ \bibnamefont {Geshkenbein}}, \bibinfo {author} {\bibfnamefont {A.~I.}\ \bibnamefont {Larkin}},\ and\ \bibinfo {author} {\bibfnamefont {V.~M.}\ \bibnamefont {Vinokur}},\ }\href {https://doi.org/10.1103/RevModPhys.66.1125} {\bibfield  {journal} {\bibinfo  {journal} {Rev. Mod. Phys.}\ }\textbf {\bibinfo {volume} {66}},\ \bibinfo {pages} {1125} (\bibinfo {year} {1994})}\BibitemShut {NoStop}%
\bibitem [{\citenamefont {Tang}(1994)}]{Tang1994TwoRepulsiveLines}%
  \BibitemOpen
  \bibfield  {author} {\bibinfo {author} {\bibfnamefont {L.-H.}\ \bibnamefont {Tang}},\ }\href {https://doi.org/https://doi.org/10.1007/BF02179451} {\bibfield  {journal} {\bibinfo  {journal} {J. Stat. Phys}\ }\textbf {\bibinfo {volume} {77}},\ \bibinfo {pages} {581} (\bibinfo {year} {1994})}\BibitemShut {NoStop}%
\bibitem [{\citenamefont {Halpin-Healy}\ and\ \citenamefont {Zhang}(1995)}]{halpin1995kinetic}%
  \BibitemOpen
  \bibfield  {author} {\bibinfo {author} {\bibfnamefont {T.}~\bibnamefont {Halpin-Healy}}\ and\ \bibinfo {author} {\bibfnamefont {Y.-C.}\ \bibnamefont {Zhang}},\ }\href {https://doi.org/https://doi.org/10.1016/0370-1573(94)00087-J} {\bibfield  {journal} {\bibinfo  {journal} {Phys. Rep.}\ }\textbf {\bibinfo {volume} {254}},\ \bibinfo {pages} {215} (\bibinfo {year} {1995})}\BibitemShut {NoStop}%
\bibitem [{\citenamefont {Balankin}\ \emph {et~al.}(2006)\citenamefont {Balankin}, \citenamefont {Paredes}, \citenamefont {Susarrey}, \citenamefont {Morales},\ and\ \citenamefont {Vacio}}]{Balankin2006PaperWetting}%
  \BibitemOpen
  \bibfield  {author} {\bibinfo {author} {\bibfnamefont {A.~S.}\ \bibnamefont {Balankin}}, \bibinfo {author} {\bibfnamefont {R.~G.}\ \bibnamefont {Paredes}}, \bibinfo {author} {\bibfnamefont {O.}~\bibnamefont {Susarrey}}, \bibinfo {author} {\bibfnamefont {D.}~\bibnamefont {Morales}},\ and\ \bibinfo {author} {\bibfnamefont {F.~C.}\ \bibnamefont {Vacio}},\ }\href {https://doi.org/10.1103/PhysRevLett.96.056101} {\bibfield  {journal} {\bibinfo  {journal} {Phys. Rev. Lett.}\ }\textbf {\bibinfo {volume} {96}},\ \bibinfo {pages} {056101} (\bibinfo {year} {2006})}\BibitemShut {NoStop}%
\bibitem [{\citenamefont {Metaxas}\ \emph {et~al.}(2010)\citenamefont {Metaxas}, \citenamefont {Stamps}, \citenamefont {Jamet}, \citenamefont {Ferr\'e}, \citenamefont {Baltz}, \citenamefont {Rodmacq},\ and\ \citenamefont {Politi}}]{MetaxasPoliti2010CoupledDomainWalls}%
  \BibitemOpen
  \bibfield  {author} {\bibinfo {author} {\bibfnamefont {P.~J.}\ \bibnamefont {Metaxas}}, \bibinfo {author} {\bibfnamefont {R.~L.}\ \bibnamefont {Stamps}}, \bibinfo {author} {\bibfnamefont {J.-P.}\ \bibnamefont {Jamet}}, \bibinfo {author} {\bibfnamefont {J.}~\bibnamefont {Ferr\'e}}, \bibinfo {author} {\bibfnamefont {V.}~\bibnamefont {Baltz}}, \bibinfo {author} {\bibfnamefont {B.}~\bibnamefont {Rodmacq}},\ and\ \bibinfo {author} {\bibfnamefont {P.}~\bibnamefont {Politi}},\ }\href {https://doi.org/10.1103/PhysRevLett.104.237206} {\bibfield  {journal} {\bibinfo  {journal} {Phys. Rev. Lett.}\ }\textbf {\bibinfo {volume} {104}},\ \bibinfo {pages} {237206} (\bibinfo {year} {2010})}\BibitemShut {NoStop}%
\bibitem [{\citenamefont {Politi}\ \emph {et~al.}(2011)\citenamefont {Politi}, \citenamefont {Metaxas}, \citenamefont {Jamet}, \citenamefont {Stamps},\ and\ \citenamefont {Ferr\'e}}]{MetaxasPoliti2011CoupledDomainWalls}%
  \BibitemOpen
  \bibfield  {author} {\bibinfo {author} {\bibfnamefont {P.}~\bibnamefont {Politi}}, \bibinfo {author} {\bibfnamefont {P.~J.}\ \bibnamefont {Metaxas}}, \bibinfo {author} {\bibfnamefont {J.-P.}\ \bibnamefont {Jamet}}, \bibinfo {author} {\bibfnamefont {R.~L.}\ \bibnamefont {Stamps}},\ and\ \bibinfo {author} {\bibfnamefont {J.}~\bibnamefont {Ferr\'e}},\ }\href {https://doi.org/10.1103/PhysRevB.84.054431} {\bibfield  {journal} {\bibinfo  {journal} {Phys. Rev. B}\ }\textbf {\bibinfo {volume} {84}},\ \bibinfo {pages} {054431} (\bibinfo {year} {2011})}\BibitemShut {NoStop}%
\bibitem [{\citenamefont {Lipowsky}(1991)}]{Lipowsky1991Membranes}%
  \BibitemOpen
  \bibfield  {author} {\bibinfo {author} {\bibfnamefont {R.}~\bibnamefont {Lipowsky}},\ }\href {https://doi.org/https://doi.org/10.1038/349475a0} {\bibfield  {journal} {\bibinfo  {journal} {Nature}\ }\textbf {\bibinfo {volume} {349}},\ \bibinfo {pages} {475} (\bibinfo {year} {1991})}\BibitemShut {NoStop}%
\bibitem [{\citenamefont {Lipowsky}(1996)}]{Lipowsky1996MembraneStickers}%
  \BibitemOpen
  \bibfield  {author} {\bibinfo {author} {\bibfnamefont {R.}~\bibnamefont {Lipowsky}},\ }\href {https://doi.org/10.1103/PhysRevLett.77.1652} {\bibfield  {journal} {\bibinfo  {journal} {Phys. Rev. Lett.}\ }\textbf {\bibinfo {volume} {77}},\ \bibinfo {pages} {1652} (\bibinfo {year} {1996})}\BibitemShut {NoStop}%
\bibitem [{Sup()}]{SupplementalMaterial}%
  \BibitemOpen
  \href@noop {} {\bibinfo  {journal} {{\color{violet}{S}upplemental {M}aterial} with discussion on single step surfaces, bubbles in the initial state and their configurational weights, co-occurrence of the two evolutions in the EWKPZ case, facilitated detachment by diamonds and half-diamonds, non-monotonic saturation width in 2EW and 2KPZ cases for finite sized systems, and dynamic exponent in the 2EW case for s=0}\ }\BibitemShut {NoStop}%
\bibitem [{\citenamefont {Krug}\ and\ \citenamefont {Tang}(1994)}]{TangKrug1994PolymerConfinedExact}%
  \BibitemOpen
\bibfield  {journal} {  }\bibfield  {author} {\bibinfo {author} {\bibfnamefont {J.}~\bibnamefont {Krug}}\ and\ \bibinfo {author} {\bibfnamefont {L.-H.}\ \bibnamefont {Tang}},\ }\href {https://doi.org/10.1103/PhysRevE.50.104} {\bibfield  {journal} {\bibinfo  {journal} {Phys. Rev. E}\ }\textbf {\bibinfo {volume} {50}},\ \bibinfo {pages} {104} (\bibinfo {year} {1994})}\BibitemShut {NoStop}%
\bibitem [{\citenamefont {Krug}(1997)}]{Krug1997AdvPhys}%
  \BibitemOpen
  \bibfield  {author} {\bibinfo {author} {\bibfnamefont {J.}~\bibnamefont {Krug}},\ }\href {https://doi.org/10.1080/00018739700101498} {\bibfield  {journal} {\bibinfo  {journal} {Adv. Phys.}\ }\textbf {\bibinfo {volume} {46}},\ \bibinfo {pages} {139} (\bibinfo {year} {1997})}\BibitemShut {NoStop}%
\bibitem [{\citenamefont {Spitzer}(1970)}]{Spitzer1970ASEP}%
  \BibitemOpen
  \bibfield  {author} {\bibinfo {author} {\bibfnamefont {F.}~\bibnamefont {Spitzer}},\ }\href {https://doi.org/https://doi.org/10.1016/0001-8708(70)90034-4} {\bibfield  {journal} {\bibinfo  {journal} {Adv. Math.}\ }\textbf {\bibinfo {volume} {5}},\ \bibinfo {pages} {246} (\bibinfo {year} {1970})}\BibitemShut {NoStop}%
\bibitem [{\citenamefont {Derrida}(1998)}]{Derrida1998ReviewASEP}%
  \BibitemOpen
  \bibfield  {author} {\bibinfo {author} {\bibfnamefont {B.}~\bibnamefont {Derrida}},\ }\href {https://doi.org/https://doi.org/10.1016/S0370-1573(98)00006-4} {\bibfield  {journal} {\bibinfo  {journal} {Phys. Rep.}\ }\textbf {\bibinfo {volume} {301}},\ \bibinfo {pages} {65} (\bibinfo {year} {1998})}\BibitemShut {NoStop}%
\bibitem [{\citenamefont {Derrida}(2007)}]{Derrida2007NonEqRev}%
  \BibitemOpen
  \bibfield  {author} {\bibinfo {author} {\bibfnamefont {B.}~\bibnamefont {Derrida}},\ }\href {https://doi.org/10.1088/1742-5468/2007/07/P07023} {\bibfield  {journal} {\bibinfo  {journal} {J. Stat. Mech.: Theory Exp.}\ }\textbf {\bibinfo {volume} {2007}}\bibinfo  {number} { (07)},\ \bibinfo {pages} {P07023}}\BibitemShut {NoStop}%
\bibitem [{\citenamefont {Blythe}\ and\ \citenamefont {Evans}(2007)}]{BlytheEvans2007SolversGuide}%
  \BibitemOpen
\bibfield  {number} {  }\bibfield  {author} {\bibinfo {author} {\bibfnamefont {R.~A.}\ \bibnamefont {Blythe}}\ and\ \bibinfo {author} {\bibfnamefont {M.~R.}\ \bibnamefont {Evans}},\ }\href {https://doi.org/10.1088/1751-8113/40/46/R01} {\bibfield  {journal} {\bibinfo  {journal} {J. Phys. A: Math. Theor.}\ }\textbf {\bibinfo {volume} {40}},\ \bibinfo {pages} {R333} (\bibinfo {year} {2007})}\BibitemShut {NoStop}%
\bibitem [{\citenamefont {Schadschneider}\ \emph {et~al.}(2010)\citenamefont {Schadschneider}, \citenamefont {Chowdhury},\ and\ \citenamefont {Nishinari}}]{SchadschneiderChowdhury2010StochasticTransport}%
  \BibitemOpen
  \bibfield  {author} {\bibinfo {author} {\bibfnamefont {A.}~\bibnamefont {Schadschneider}}, \bibinfo {author} {\bibfnamefont {D.}~\bibnamefont {Chowdhury}},\ and\ \bibinfo {author} {\bibfnamefont {K.}~\bibnamefont {Nishinari}},\ }\href {https://doi.org/https://doi.org/10.1016/C2009-0-16900-3} {\emph {\bibinfo {title} {Stochastic Transport in Complex Systems: From Molecules to Vehicles}}}\ (\bibinfo  {publisher} {Elsevier},\ \bibinfo {year} {2010})\BibitemShut {NoStop}%
\bibitem [{\citenamefont {Mallick}(2015)}]{Mallick2015ReviewASEP}%
  \BibitemOpen
  \bibfield  {author} {\bibinfo {author} {\bibfnamefont {K.}~\bibnamefont {Mallick}},\ }\href {https://doi.org/https://doi.org/10.1016/j.physa.2014.07.046} {\bibfield  {journal} {\bibinfo  {journal} {Physica A}\ }\textbf {\bibinfo {volume} {418}},\ \bibinfo {pages} {17} (\bibinfo {year} {2015})}\BibitemShut {NoStop}%
\bibitem [{\citenamefont {Odijk}(1983)}]{Odijk1983ConfinedPolymers}%
  \BibitemOpen
  \bibfield  {author} {\bibinfo {author} {\bibfnamefont {T.}~\bibnamefont {Odijk}},\ }\href {https://doi.org/10.1021/ma00242a015} {\bibfield  {journal} {\bibinfo  {journal} {Macromolecules}\ }\textbf {\bibinfo {volume} {16}},\ \bibinfo {pages} {1340} (\bibinfo {year} {1983})}\BibitemShut {NoStop}%
\bibitem [{\citenamefont {Reisner}\ \emph {et~al.}(2005)\citenamefont {Reisner}, \citenamefont {Morton}, \citenamefont {Riehn}, \citenamefont {Wang}, \citenamefont {Yu}, \citenamefont {Rosen}, \citenamefont {Sturm}, \citenamefont {Chou}, \citenamefont {Frey},\ and\ \citenamefont {Austin}}]{ReisnerFrey2005DNANanochannels}%
  \BibitemOpen
  \bibfield  {author} {\bibinfo {author} {\bibfnamefont {W.}~\bibnamefont {Reisner}}, \bibinfo {author} {\bibfnamefont {K.~J.}\ \bibnamefont {Morton}}, \bibinfo {author} {\bibfnamefont {R.}~\bibnamefont {Riehn}}, \bibinfo {author} {\bibfnamefont {Y.~M.}\ \bibnamefont {Wang}}, \bibinfo {author} {\bibfnamefont {Z.}~\bibnamefont {Yu}}, \bibinfo {author} {\bibfnamefont {M.}~\bibnamefont {Rosen}}, \bibinfo {author} {\bibfnamefont {J.~C.}\ \bibnamefont {Sturm}}, \bibinfo {author} {\bibfnamefont {S.~Y.}\ \bibnamefont {Chou}}, \bibinfo {author} {\bibfnamefont {E.}~\bibnamefont {Frey}},\ and\ \bibinfo {author} {\bibfnamefont {R.~H.}\ \bibnamefont {Austin}},\ }\href {https://doi.org/10.1103/PhysRevLett.94.196101} {\bibfield  {journal} {\bibinfo  {journal} {Phys. Rev. Lett.}\ }\textbf {\bibinfo {volume} {94}},\ \bibinfo {pages} {196101} (\bibinfo {year} {2005})}\BibitemShut {NoStop}%
\bibitem [{\citenamefont {Yadav}\ \emph {et~al.}(2020)\citenamefont {Yadav}, \citenamefont {Rosencrans}, \citenamefont {Basak}, \citenamefont {van Kan},\ and\ \citenamefont {van~der Maarel}}]{Indresh2020Nanochanneldynamics}%
  \BibitemOpen
  \bibfield  {author} {\bibinfo {author} {\bibfnamefont {I.}~\bibnamefont {Yadav}}, \bibinfo {author} {\bibfnamefont {W.}~\bibnamefont {Rosencrans}}, \bibinfo {author} {\bibfnamefont {R.}~\bibnamefont {Basak}}, \bibinfo {author} {\bibfnamefont {J.~A.}\ \bibnamefont {van Kan}},\ and\ \bibinfo {author} {\bibfnamefont {J.~R.~C.}\ \bibnamefont {van~der Maarel}},\ }\href {https://doi.org/10.1103/PhysRevResearch.2.013294} {\bibfield  {journal} {\bibinfo  {journal} {Phys. Rev. Res.}\ }\textbf {\bibinfo {volume} {2}},\ \bibinfo {pages} {013294} (\bibinfo {year} {2020})}\BibitemShut {NoStop}%
\bibitem [{\citenamefont {Yadav}\ \emph {et~al.}(2024)\citenamefont {Yadav}, \citenamefont {Basak}, \citenamefont {van Kan},\ and\ \citenamefont {van~der Maarel}}]{Indresh2024Nanochanneldynamics}%
  \BibitemOpen
  \bibfield  {author} {\bibinfo {author} {\bibfnamefont {I.}~\bibnamefont {Yadav}}, \bibinfo {author} {\bibfnamefont {R.}~\bibnamefont {Basak}}, \bibinfo {author} {\bibfnamefont {J.~A.}\ \bibnamefont {van Kan}},\ and\ \bibinfo {author} {\bibfnamefont {J.~R.}\ \bibnamefont {van~der Maarel}},\ }\href {https://doi.org/10.1209/0295-5075/ad7dad} {\bibfield  {journal} {\bibinfo  {journal} {Europhys. Lett.}\ }\textbf {\bibinfo {volume} {148}},\ \bibinfo {pages} {17002} (\bibinfo {year} {2024})}\BibitemShut {NoStop}%
\bibitem [{\citenamefont {Balducci}\ \emph {et~al.}(2007)\citenamefont {Balducci}, \citenamefont {Hsieh},\ and\ \citenamefont {Doyle}}]{Doyle2007DNASlitlike}%
  \BibitemOpen
  \bibfield  {author} {\bibinfo {author} {\bibfnamefont {A.}~\bibnamefont {Balducci}}, \bibinfo {author} {\bibfnamefont {C.-C.}\ \bibnamefont {Hsieh}},\ and\ \bibinfo {author} {\bibfnamefont {P.~S.}\ \bibnamefont {Doyle}},\ }\href {https://doi.org/10.1103/PhysRevLett.99.238102} {\bibfield  {journal} {\bibinfo  {journal} {Phys. Rev. Lett.}\ }\textbf {\bibinfo {volume} {99}},\ \bibinfo {pages} {238102} (\bibinfo {year} {2007})}\BibitemShut {NoStop}%
\bibitem [{\citenamefont {Tang}\ and\ \citenamefont {Doyle}(2007)}]{TangDoyle2007Nanochannels}%
  \BibitemOpen
  \bibfield  {author} {\bibinfo {author} {\bibfnamefont {J.}~\bibnamefont {Tang}}\ and\ \bibinfo {author} {\bibfnamefont {P.~S.}\ \bibnamefont {Doyle}},\ }\href {https://doi.org/10.1063/1.2745650} {\bibfield  {journal} {\bibinfo  {journal} {Appl. Phys. Lett.}\ }\textbf {\bibinfo {volume} {90}} (\bibinfo {year} {2007})}\BibitemShut {NoStop}%
\bibitem [{\citenamefont {Balducci}\ \emph {et~al.}(2008)\citenamefont {Balducci}, \citenamefont {Tang},\ and\ \citenamefont {Doyle}}]{BalducciDoyle2008Nanoslit}%
  \BibitemOpen
  \bibfield  {author} {\bibinfo {author} {\bibfnamefont {A.~G.}\ \bibnamefont {Balducci}}, \bibinfo {author} {\bibfnamefont {J.}~\bibnamefont {Tang}},\ and\ \bibinfo {author} {\bibfnamefont {P.~S.}\ \bibnamefont {Doyle}},\ }\href {https://doi.org/https://pubs.acs.org/doi/10.1021/ma8015344} {\bibfield  {journal} {\bibinfo  {journal} {Macromolecules}\ }\textbf {\bibinfo {volume} {41}},\ \bibinfo {pages} {9914} (\bibinfo {year} {2008})}\BibitemShut {NoStop}%
\bibitem [{\citenamefont {Jones}\ \emph {et~al.}(2013)\citenamefont {Jones}, \citenamefont {van~der Maarel},\ and\ \citenamefont {Doyle}}]{Doyle2013IntrachainDNASlitlike}%
  \BibitemOpen
  \bibfield  {author} {\bibinfo {author} {\bibfnamefont {J.~J.}\ \bibnamefont {Jones}}, \bibinfo {author} {\bibfnamefont {J.~R.~C.}\ \bibnamefont {van~der Maarel}},\ and\ \bibinfo {author} {\bibfnamefont {P.~S.}\ \bibnamefont {Doyle}},\ }\href {https://doi.org/10.1103/PhysRevLett.110.068101} {\bibfield  {journal} {\bibinfo  {journal} {Phys. Rev. Lett.}\ }\textbf {\bibinfo {volume} {110}},\ \bibinfo {pages} {068101} (\bibinfo {year} {2013})}\BibitemShut {NoStop}%
\bibitem [{\citenamefont {Maier}\ and\ \citenamefont {R\"adler}(1999)}]{MaierRadler1999DNALipids}%
  \BibitemOpen
  \bibfield  {author} {\bibinfo {author} {\bibfnamefont {B.}~\bibnamefont {Maier}}\ and\ \bibinfo {author} {\bibfnamefont {J.~O.}\ \bibnamefont {R\"adler}},\ }\href {https://doi.org/10.1103/PhysRevLett.82.1911} {\bibfield  {journal} {\bibinfo  {journal} {Phys. Rev. Lett.}\ }\textbf {\bibinfo {volume} {82}},\ \bibinfo {pages} {1911} (\bibinfo {year} {1999})}\BibitemShut {NoStop}%
\bibitem [{\citenamefont {Barab\'asi}(1992)}]{Barabasi1992CoupledInterfaces}%
  \BibitemOpen
  \bibfield  {author} {\bibinfo {author} {\bibfnamefont {A.-L.}\ \bibnamefont {Barab\'asi}},\ }\href {https://doi.org/10.1103/PhysRevA.46.R2977} {\bibfield  {journal} {\bibinfo  {journal} {Phys. Rev. A}\ }\textbf {\bibinfo {volume} {46}},\ \bibinfo {pages} {R2977} (\bibinfo {year} {1992})}\BibitemShut {NoStop}%
\bibitem [{\citenamefont {Barab{\'a}si}(1993)}]{Barabasi1993SurfactantCoupledInterfaces}%
  \BibitemOpen
  \bibfield  {author} {\bibinfo {author} {\bibfnamefont {A.-L.}\ \bibnamefont {Barab{\'a}si}},\ }\href {https://doi.org/10.1103/PhysRevLett.70.4102} {\bibfield  {journal} {\bibinfo  {journal} {Phys. Rev. Lett.}\ }\textbf {\bibinfo {volume} {70}},\ \bibinfo {pages} {4102} (\bibinfo {year} {1993})}\BibitemShut {NoStop}%
\bibitem [{\citenamefont {Erta\ifmmode~\mbox{\c{s}}\else \c{s}\fi{}}\ and\ \citenamefont {Kardar}(1992)}]{ErtasKardar1992DirectedLines}%
  \BibitemOpen
  \bibfield  {author} {\bibinfo {author} {\bibfnamefont {D.}~\bibnamefont {Erta\ifmmode~\mbox{\c{s}}\else \c{s}\fi{}}}\ and\ \bibinfo {author} {\bibfnamefont {M.}~\bibnamefont {Kardar}},\ }\href {https://doi.org/10.1103/PhysRevLett.69.929} {\bibfield  {journal} {\bibinfo  {journal} {Phys. Rev. Lett.}\ }\textbf {\bibinfo {volume} {69}},\ \bibinfo {pages} {929} (\bibinfo {year} {1992})}\BibitemShut {NoStop}%
\bibitem [{\citenamefont {Erta\ifmmode~\mbox{\c{s}}\else \c{s}\fi{}}\ and\ \citenamefont {Kardar}(1993)}]{ErtasKardar1993DriftingPolymers}%
  \BibitemOpen
  \bibfield  {author} {\bibinfo {author} {\bibfnamefont {D.}~\bibnamefont {Erta\ifmmode~\mbox{\c{s}}\else \c{s}\fi{}}}\ and\ \bibinfo {author} {\bibfnamefont {M.}~\bibnamefont {Kardar}},\ }\href {https://doi.org/10.1103/PhysRevE.48.1228} {\bibfield  {journal} {\bibinfo  {journal} {Phys. Rev. E}\ }\textbf {\bibinfo {volume} {48}},\ \bibinfo {pages} {1228} (\bibinfo {year} {1993})}\BibitemShut {NoStop}%
\bibitem [{\citenamefont {Majumdar}\ and\ \citenamefont {Das}(2005)}]{SatyaDibyendu2005PersistenceCoupledInterfaces}%
  \BibitemOpen
  \bibfield  {author} {\bibinfo {author} {\bibfnamefont {S.~N.}\ \bibnamefont {Majumdar}}\ and\ \bibinfo {author} {\bibfnamefont {D.}~\bibnamefont {Das}},\ }\href {https://doi.org/10.1103/PhysRevE.71.036129} {\bibfield  {journal} {\bibinfo  {journal} {Phys. Rev. E}\ }\textbf {\bibinfo {volume} {71}},\ \bibinfo {pages} {036129} (\bibinfo {year} {2005})}\BibitemShut {NoStop}%
\bibitem [{\citenamefont {Juntunen}\ \emph {et~al.}(2007)\citenamefont {Juntunen}, \citenamefont {Pulkkinen},\ and\ \citenamefont {Merikoski}}]{Juntunen2007Interfaces}%
  \BibitemOpen
  \bibfield  {author} {\bibinfo {author} {\bibfnamefont {J.}~\bibnamefont {Juntunen}}, \bibinfo {author} {\bibfnamefont {O.}~\bibnamefont {Pulkkinen}},\ and\ \bibinfo {author} {\bibfnamefont {J.}~\bibnamefont {Merikoski}},\ }\href {https://doi.org/10.1103/PhysRevE.76.041607} {\bibfield  {journal} {\bibinfo  {journal} {Phys. Rev. E}\ }\textbf {\bibinfo {volume} {76}},\ \bibinfo {pages} {041607} (\bibinfo {year} {2007})}\BibitemShut {NoStop}%
\bibitem [{\citenamefont {Ferrari}\ \emph {et~al.}(2013)\citenamefont {Ferrari}, \citenamefont {Sasamoto},\ and\ \citenamefont {Spohn}}]{FerrariSpohn2013CoupledKPZ1D}%
  \BibitemOpen
  \bibfield  {author} {\bibinfo {author} {\bibfnamefont {P.~L.}\ \bibnamefont {Ferrari}}, \bibinfo {author} {\bibfnamefont {T.}~\bibnamefont {Sasamoto}},\ and\ \bibinfo {author} {\bibfnamefont {H.}~\bibnamefont {Spohn}},\ }\href {https://doi.org/https://doi.org/10.1007/s10955-013-0842-5} {\bibfield  {journal} {\bibinfo  {journal} {J. Stat. Phys.}\ }\textbf {\bibinfo {volume} {153}},\ \bibinfo {pages} {377} (\bibinfo {year} {2013})}\BibitemShut {NoStop}%
\bibitem [{\citenamefont {Mendl}\ and\ \citenamefont {Spohn}(2013)}]{MendlSpohn2013nlf-hydrodynamics}%
  \BibitemOpen
  \bibfield  {author} {\bibinfo {author} {\bibfnamefont {C.~B.}\ \bibnamefont {Mendl}}\ and\ \bibinfo {author} {\bibfnamefont {H.}~\bibnamefont {Spohn}},\ }\href {https://doi.org/10.1103/PhysRevLett.111.230601} {\bibfield  {journal} {\bibinfo  {journal} {Phys. Rev. Lett.}\ }\textbf {\bibinfo {volume} {111}},\ \bibinfo {pages} {230601} (\bibinfo {year} {2013})}\BibitemShut {NoStop}%
\bibitem [{\citenamefont {Spohn}(2014)}]{Spohn2014AnharmonicChains}%
  \BibitemOpen
  \bibfield  {author} {\bibinfo {author} {\bibfnamefont {H.}~\bibnamefont {Spohn}},\ }\href {https://doi.org/https://doi.org/10.1007/s10955-014-0933-y} {\bibfield  {journal} {\bibinfo  {journal} {J. Stat. Phys.}\ }\textbf {\bibinfo {volume} {154}},\ \bibinfo {pages} {1191} (\bibinfo {year} {2014})}\BibitemShut {NoStop}%
\bibitem [{\citenamefont {Spohn}\ and\ \citenamefont {Stoltz}(2015)}]{SpohnStoltz2015coupledKPZ}%
  \BibitemOpen
  \bibfield  {author} {\bibinfo {author} {\bibfnamefont {H.}~\bibnamefont {Spohn}}\ and\ \bibinfo {author} {\bibfnamefont {G.}~\bibnamefont {Stoltz}},\ }\href {https://doi.org/https://doi.org/10.1007/s10955-015-1214-0} {\bibfield  {journal} {\bibinfo  {journal} {J. Stat. Phys.}\ }\textbf {\bibinfo {volume} {160}},\ \bibinfo {pages} {861} (\bibinfo {year} {2015})}\BibitemShut {NoStop}%
\bibitem [{\citenamefont {Sch\"utz}\ and\ \citenamefont {Wehefritz-Kaufmann}(2017)}]{SchutzKaufmann2017DdimDirectedPolymers}%
  \BibitemOpen
  \bibfield  {author} {\bibinfo {author} {\bibfnamefont {G.~M.}\ \bibnamefont {Sch\"utz}}\ and\ \bibinfo {author} {\bibfnamefont {B.}~\bibnamefont {Wehefritz-Kaufmann}},\ }\href {https://doi.org/10.1103/PhysRevE.96.032119} {\bibfield  {journal} {\bibinfo  {journal} {Phys. Rev. E}\ }\textbf {\bibinfo {volume} {96}},\ \bibinfo {pages} {032119} (\bibinfo {year} {2017})}\BibitemShut {NoStop}%
\bibitem [{\citenamefont {Bernardin}\ \emph {et~al.}(2021)\citenamefont {Bernardin}, \citenamefont {Funaki},\ and\ \citenamefont {Sethuraman}}]{Bernardin2021coupledKPZBurgers}%
  \BibitemOpen
  \bibfield  {author} {\bibinfo {author} {\bibfnamefont {C.}~\bibnamefont {Bernardin}}, \bibinfo {author} {\bibfnamefont {T.}~\bibnamefont {Funaki}},\ and\ \bibinfo {author} {\bibfnamefont {S.}~\bibnamefont {Sethuraman}},\ }\href {https://doi.org/10.1214/20-AAP1639} {\bibfield  {journal} {\bibinfo  {journal} {Ann. Appl. Probab.}\ }\textbf {\bibinfo {volume} {31}},\ \bibinfo {pages} {1966} (\bibinfo {year} {2021})}\BibitemShut {NoStop}%
\bibitem [{\citenamefont {De~Nardis}\ \emph {et~al.}(2023)\citenamefont {De~Nardis}, \citenamefont {Gopalakrishnan},\ and\ \citenamefont {Vasseur}}]{DeNardisGopalakrishnan2023SpinChainsKPZ}%
  \BibitemOpen
  \bibfield  {author} {\bibinfo {author} {\bibfnamefont {J.}~\bibnamefont {De~Nardis}}, \bibinfo {author} {\bibfnamefont {S.}~\bibnamefont {Gopalakrishnan}},\ and\ \bibinfo {author} {\bibfnamefont {R.}~\bibnamefont {Vasseur}},\ }\href {https://doi.org/10.1103/PhysRevLett.131.197102} {\bibfield  {journal} {\bibinfo  {journal} {Phys. Rev. Lett.}\ }\textbf {\bibinfo {volume} {131}},\ \bibinfo {pages} {197102} (\bibinfo {year} {2023})}\BibitemShut {NoStop}%
\bibitem [{\citenamefont {Roy}\ \emph {et~al.}(2024)\citenamefont {Roy}, \citenamefont {Dhar}, \citenamefont {Khanin}, \citenamefont {Kulkarni},\ and\ \citenamefont {Spohn}}]{RoyDharSpohn2024coupledburgers}%
  \BibitemOpen
  \bibfield  {author} {\bibinfo {author} {\bibfnamefont {D.}~\bibnamefont {Roy}}, \bibinfo {author} {\bibfnamefont {A.}~\bibnamefont {Dhar}}, \bibinfo {author} {\bibfnamefont {K.}~\bibnamefont {Khanin}}, \bibinfo {author} {\bibfnamefont {M.}~\bibnamefont {Kulkarni}},\ and\ \bibinfo {author} {\bibfnamefont {H.}~\bibnamefont {Spohn}},\ }\href {https://doi.org/10.1088/1742-5468/ad3196} {\bibfield  {journal} {\bibinfo  {journal} {J. Stat. Mech.: Theory Exp.}\ }\textbf {\bibinfo {volume} {2024}}\bibinfo  {number} { (3)},\ \bibinfo {pages} {033209}}\BibitemShut {NoStop}%
\bibitem [{\citenamefont {Roy}\ \emph {et~al.}(2025)\citenamefont {Roy}, \citenamefont {Dhar}, \citenamefont {Kulkarni},\ and\ \citenamefont {Spohn}}]{RoyDharSpohn2025coupledKPZ}%
  \BibitemOpen
\bibfield  {number} {  }\bibfield  {author} {\bibinfo {author} {\bibfnamefont {D.}~\bibnamefont {Roy}}, \bibinfo {author} {\bibfnamefont {A.}~\bibnamefont {Dhar}}, \bibinfo {author} {\bibfnamefont {M.}~\bibnamefont {Kulkarni}},\ and\ \bibinfo {author} {\bibfnamefont {H.}~\bibnamefont {Spohn}},\ }\href {https://arxiv.org/abs/2504.04162} {\bibfield  {journal} {\bibinfo  {journal} {arXiv:2504.04162}\ } (\bibinfo {year} {2025})}\BibitemShut {NoStop}%
\bibitem [{\citenamefont {Lahiri}\ and\ \citenamefont {Ramaswamy}(1997)}]{LR1997Sedimentation}%
  \BibitemOpen
  \bibfield  {author} {\bibinfo {author} {\bibfnamefont {R.}~\bibnamefont {Lahiri}}\ and\ \bibinfo {author} {\bibfnamefont {S.}~\bibnamefont {Ramaswamy}},\ }\href {https://doi.org/10.1103/PhysRevLett.79.1150} {\bibfield  {journal} {\bibinfo  {journal} {Phys. Rev. Lett.}\ }\textbf {\bibinfo {volume} {79}},\ \bibinfo {pages} {1150} (\bibinfo {year} {1997})}\BibitemShut {NoStop}%
\bibitem [{\citenamefont {Lahiri}\ \emph {et~al.}(2000)\citenamefont {Lahiri}, \citenamefont {Barma},\ and\ \citenamefont {Ramaswamy}}]{LBR2000SPS}%
  \BibitemOpen
  \bibfield  {author} {\bibinfo {author} {\bibfnamefont {R.}~\bibnamefont {Lahiri}}, \bibinfo {author} {\bibfnamefont {M.}~\bibnamefont {Barma}},\ and\ \bibinfo {author} {\bibfnamefont {S.}~\bibnamefont {Ramaswamy}},\ }\href {https://doi.org/10.1103/PhysRevE.61.1648} {\bibfield  {journal} {\bibinfo  {journal} {Phys. Rev. E}\ }\textbf {\bibinfo {volume} {61}},\ \bibinfo {pages} {1648} (\bibinfo {year} {2000})}\BibitemShut {NoStop}%
\bibitem [{\citenamefont {Arndt}\ \emph {et~al.}(1998)\citenamefont {Arndt}, \citenamefont {Heinzel},\ and\ \citenamefont {Rittenberg}}]{ArndtHeinzelRittenberg1998AHRoriginal}%
  \BibitemOpen
  \bibfield  {author} {\bibinfo {author} {\bibfnamefont {P.~F.}\ \bibnamefont {Arndt}}, \bibinfo {author} {\bibfnamefont {T.}~\bibnamefont {Heinzel}},\ and\ \bibinfo {author} {\bibfnamefont {V.}~\bibnamefont {Rittenberg}},\ }\href {https://iopscience.iop.org/article/10.1088/0305-4470/31/2/001} {\bibfield  {journal} {\bibinfo  {journal} {J. Phys. A: Math. Gen.}\ }\textbf {\bibinfo {volume} {31}},\ \bibinfo {pages} {L45} (\bibinfo {year} {1998})}\BibitemShut {NoStop}%
\bibitem [{\citenamefont {Rajewsky}\ \emph {et~al.}(2000)\citenamefont {Rajewsky}, \citenamefont {Sasamoto},\ and\ \citenamefont {Speer}}]{Rajewsky2000AHRRing}%
  \BibitemOpen
  \bibfield  {author} {\bibinfo {author} {\bibfnamefont {N.}~\bibnamefont {Rajewsky}}, \bibinfo {author} {\bibfnamefont {T.}~\bibnamefont {Sasamoto}},\ and\ \bibinfo {author} {\bibfnamefont {E.}~\bibnamefont {Speer}},\ }\href {https://doi.org/https://doi.org/10.1016/S0378-4371(99)00537-3} {\bibfield  {journal} {\bibinfo  {journal} {Physica A}\ }\textbf {\bibinfo {volume} {279}},\ \bibinfo {pages} {123} (\bibinfo {year} {2000})}\BibitemShut {NoStop}%
\bibitem [{\citenamefont {Popkov}\ \emph {et~al.}(2014)\citenamefont {Popkov}, \citenamefont {Schmidt},\ and\ \citenamefont {Sch{\"u}tz}}]{Popkov2014SuperdiffusiveModes}%
  \BibitemOpen
  \bibfield  {author} {\bibinfo {author} {\bibfnamefont {V.}~\bibnamefont {Popkov}}, \bibinfo {author} {\bibfnamefont {J.}~\bibnamefont {Schmidt}},\ and\ \bibinfo {author} {\bibfnamefont {G.}~\bibnamefont {Sch{\"u}tz}},\ }\href {https://doi.org/10.1103/PhysRevLett.112.200602} {\bibfield  {journal} {\bibinfo  {journal} {Phys. Rev. Lett.}\ }\textbf {\bibinfo {volume} {112}},\ \bibinfo {pages} {200602} (\bibinfo {year} {2014})}\BibitemShut {NoStop}%
\bibitem [{\citenamefont {Popkov}\ \emph {et~al.}(2015)\citenamefont {Popkov}, \citenamefont {Schadschneider}, \citenamefont {Schmidt},\ and\ \citenamefont {Sch{\"u}tz}}]{Popkov2015Fibonacci-universality}%
  \BibitemOpen
  \bibfield  {author} {\bibinfo {author} {\bibfnamefont {V.}~\bibnamefont {Popkov}}, \bibinfo {author} {\bibfnamefont {A.}~\bibnamefont {Schadschneider}}, \bibinfo {author} {\bibfnamefont {J.}~\bibnamefont {Schmidt}},\ and\ \bibinfo {author} {\bibfnamefont {G.~M.}\ \bibnamefont {Sch{\"u}tz}},\ }\href {https://doi.org/10.1073/pnas.1512261112} {\bibfield  {journal} {\bibinfo  {journal} {Proc. Natl. Acad. Sci. U.S.A.}\ }\textbf {\bibinfo {volume} {112}},\ \bibinfo {pages} {12645} (\bibinfo {year} {2015})}\BibitemShut {NoStop}%
\bibitem [{\citenamefont {Chakraborty}\ \emph {et~al.}(2017{\natexlab{a}})\citenamefont {Chakraborty}, \citenamefont {Chatterjee},\ and\ \citenamefont {Barma}}]{Chakraborty2017LH-statics}%
  \BibitemOpen
  \bibfield  {author} {\bibinfo {author} {\bibfnamefont {S.}~\bibnamefont {Chakraborty}}, \bibinfo {author} {\bibfnamefont {S.}~\bibnamefont {Chatterjee}},\ and\ \bibinfo {author} {\bibfnamefont {M.}~\bibnamefont {Barma}},\ }\href {https://doi.org/10.1103/PhysRevE.96.022127} {\bibfield  {journal} {\bibinfo  {journal} {Phys. Rev. E}\ }\textbf {\bibinfo {volume} {96}},\ \bibinfo {pages} {022127} (\bibinfo {year} {2017}{\natexlab{a}})}\BibitemShut {NoStop}%
\bibitem [{\citenamefont {Chakraborty}\ \emph {et~al.}(2017{\natexlab{b}})\citenamefont {Chakraborty}, \citenamefont {Chatterjee},\ and\ \citenamefont {Barma}}]{Chakraborty2017LH-dynamics}%
  \BibitemOpen
  \bibfield  {author} {\bibinfo {author} {\bibfnamefont {S.}~\bibnamefont {Chakraborty}}, \bibinfo {author} {\bibfnamefont {S.}~\bibnamefont {Chatterjee}},\ and\ \bibinfo {author} {\bibfnamefont {M.}~\bibnamefont {Barma}},\ }\href {https://doi.org/10.1103/PhysRevE.96.022128} {\bibfield  {journal} {\bibinfo  {journal} {Phys. Rev. E}\ }\textbf {\bibinfo {volume} {96}},\ \bibinfo {pages} {022128} (\bibinfo {year} {2017}{\natexlab{b}})}\BibitemShut {NoStop}%
\bibitem [{\citenamefont {Mahapatra}\ \emph {et~al.}(2020)\citenamefont {Mahapatra}, \citenamefont {Ramola},\ and\ \citenamefont {Barma}}]{Mahapatra2020lightHeavy}%
  \BibitemOpen
  \bibfield  {author} {\bibinfo {author} {\bibfnamefont {S.}~\bibnamefont {Mahapatra}}, \bibinfo {author} {\bibfnamefont {K.}~\bibnamefont {Ramola}},\ and\ \bibinfo {author} {\bibfnamefont {M.}~\bibnamefont {Barma}},\ }\href {https://doi.org/10.1103/PhysRevResearch.2.043279} {\bibfield  {journal} {\bibinfo  {journal} {Phys. Rev. Res.}\ }\textbf {\bibinfo {volume} {2}},\ \bibinfo {pages} {043279} (\bibinfo {year} {2020})}\BibitemShut {NoStop}%
\bibitem [{\citenamefont {Chen}\ \emph {et~al.}(2018)\citenamefont {Chen}, \citenamefont {de~Gier}, \citenamefont {Hiki},\ and\ \citenamefont {Sasamoto}}]{Sasamoto2018Two-Species}%
  \BibitemOpen
  \bibfield  {author} {\bibinfo {author} {\bibfnamefont {Z.}~\bibnamefont {Chen}}, \bibinfo {author} {\bibfnamefont {J.}~\bibnamefont {de~Gier}}, \bibinfo {author} {\bibfnamefont {I.}~\bibnamefont {Hiki}},\ and\ \bibinfo {author} {\bibfnamefont {T.}~\bibnamefont {Sasamoto}},\ }\href {https://doi.org/10.1103/PhysRevLett.120.240601} {\bibfield  {journal} {\bibinfo  {journal} {Phys. Rev. Lett.}\ }\textbf {\bibinfo {volume} {120}},\ \bibinfo {pages} {240601} (\bibinfo {year} {2018})}\BibitemShut {NoStop}%
\bibitem [{\citenamefont {Chen}\ \emph {et~al.}(2022)\citenamefont {Chen}, \citenamefont {de~Gier}, \citenamefont {Hiki}, \citenamefont {Sasamoto},\ and\ \citenamefont {Usui}}]{Sasamoto2022Two-SpeciesAHR}%
  \BibitemOpen
  \bibfield  {author} {\bibinfo {author} {\bibfnamefont {Z.}~\bibnamefont {Chen}}, \bibinfo {author} {\bibfnamefont {J.}~\bibnamefont {de~Gier}}, \bibinfo {author} {\bibfnamefont {I.}~\bibnamefont {Hiki}}, \bibinfo {author} {\bibfnamefont {T.}~\bibnamefont {Sasamoto}},\ and\ \bibinfo {author} {\bibfnamefont {M.}~\bibnamefont {Usui}},\ }\href {https://doi.org/https://doi.org/10.1007/s00220-022-04408-8} {\bibfield  {journal} {\bibinfo  {journal} {Commun. Math. Phys}\ }\textbf {\bibinfo {volume} {395}},\ \bibinfo {pages} {59} (\bibinfo {year} {2022})}\BibitemShut {NoStop}%
\bibitem [{\citenamefont {Schmidt}\ \emph {et~al.}(2021)\citenamefont {Schmidt}, \citenamefont {Sch{\"u}tz},\ and\ \citenamefont {van Beijeren}}]{SchmidtSchutzvanBeijeren2021ThreeLane}%
  \BibitemOpen
  \bibfield  {author} {\bibinfo {author} {\bibfnamefont {J.}~\bibnamefont {Schmidt}}, \bibinfo {author} {\bibfnamefont {G.~M.}\ \bibnamefont {Sch{\"u}tz}},\ and\ \bibinfo {author} {\bibfnamefont {H.}~\bibnamefont {van Beijeren}},\ }\href {https://doi.org/https://doi.org/10.1007/s10955-021-02709-1} {\bibfield  {journal} {\bibinfo  {journal} {J. Stat. Phys.}\ }\textbf {\bibinfo {volume} {183}},\ \bibinfo {pages} {8} (\bibinfo {year} {2021})}\BibitemShut {NoStop}%
\bibitem [{\citenamefont {Dolai}\ \emph {et~al.}(2024)\citenamefont {Dolai}, \citenamefont {Simha},\ and\ \citenamefont {Basu}}]{DolaiSimhaAbhikBasu2024CoupledASEP}%
  \BibitemOpen
  \bibfield  {author} {\bibinfo {author} {\bibfnamefont {P.}~\bibnamefont {Dolai}}, \bibinfo {author} {\bibfnamefont {A.}~\bibnamefont {Simha}},\ and\ \bibinfo {author} {\bibfnamefont {A.}~\bibnamefont {Basu}},\ }\href {https://doi.org/10.1103/PhysRevE.109.064122} {\bibfield  {journal} {\bibinfo  {journal} {Phys. Rev. E}\ }\textbf {\bibinfo {volume} {109}},\ \bibinfo {pages} {064122} (\bibinfo {year} {2024})}\BibitemShut {NoStop}%
\bibitem [{\citenamefont {Cannizzaro}\ \emph {et~al.}(2025)\citenamefont {Cannizzaro}, \citenamefont {Gon{\c{c}}alves}, \citenamefont {Misturini},\ and\ \citenamefont {Occelli}}]{CannizzaroOccelli2025FromABCtoKPZ}%
  \BibitemOpen
  \bibfield  {author} {\bibinfo {author} {\bibfnamefont {G.}~\bibnamefont {Cannizzaro}}, \bibinfo {author} {\bibfnamefont {P.}~\bibnamefont {Gon{\c{c}}alves}}, \bibinfo {author} {\bibfnamefont {R.}~\bibnamefont {Misturini}},\ and\ \bibinfo {author} {\bibfnamefont {A.}~\bibnamefont {Occelli}},\ }\href {https://doi.org/https://doi.org/10.1007/s00440-024-01314-z} {\bibfield  {journal} {\bibinfo  {journal} {Probab. Theory Relat. Fields}\ }\textbf {\bibinfo {volume} {191}},\ \bibinfo {pages} {361} (\bibinfo {year} {2025})}\BibitemShut {NoStop}%
\bibitem [{\citenamefont {Ferrari}\ and\ \citenamefont {Gernholt}(2025)}]{FerrariGernholt2025DecouplingDecayTasep}%
  \BibitemOpen
  \bibfield  {author} {\bibinfo {author} {\bibfnamefont {P.~L.}\ \bibnamefont {Ferrari}}\ and\ \bibinfo {author} {\bibfnamefont {S.}~\bibnamefont {Gernholt}},\ }\href {https://arxiv.org/abs/2504.00765} {\bibfield  {journal} {\bibinfo  {journal} {arXiv:2504.00765}\ } (\bibinfo {year} {2025})}\BibitemShut {NoStop}%
\bibitem [{\citenamefont {Prakash}\ \emph {et~al.}(2025)\citenamefont {Prakash}, \citenamefont {Barma},\ and\ \citenamefont {Ramola}}]{PrakashKabir2025scaledLH}%
  \BibitemOpen
  \bibfield  {author} {\bibinfo {author} {\bibfnamefont {S.}~\bibnamefont {Prakash}}, \bibinfo {author} {\bibfnamefont {M.}~\bibnamefont {Barma}},\ and\ \bibinfo {author} {\bibfnamefont {K.}~\bibnamefont {Ramola}},\ }\href {https://arxiv.org/abs/2503.16103} {\bibfield  {journal} {\bibinfo  {journal} {arXiv:2503.16103}\ } (\bibinfo {year} {2025})}\BibitemShut {NoStop}%
\bibitem [{\citenamefont {Sch{\"u}tz}\ \emph {et~al.}(1996)\citenamefont {Sch{\"u}tz}, \citenamefont {Ramaswamy},\ and\ \citenamefont {Barma}}]{Schutz1996Pairwise}%
  \BibitemOpen
  \bibfield  {author} {\bibinfo {author} {\bibfnamefont {G.~M.}\ \bibnamefont {Sch{\"u}tz}}, \bibinfo {author} {\bibfnamefont {R.}~\bibnamefont {Ramaswamy}},\ and\ \bibinfo {author} {\bibfnamefont {M.}~\bibnamefont {Barma}},\ }\href {https://doi.org/10.1088/0305-4470/29/4/011} {\bibfield  {journal} {\bibinfo  {journal} {J. Phys. A: Math. Gen.}\ }\textbf {\bibinfo {volume} {29}},\ \bibinfo {pages} {837} (\bibinfo {year} {1996})}\BibitemShut {NoStop}%
\bibitem [{\citenamefont {Tripathy}\ and\ \citenamefont {Barma}(1997)}]{TripathyBarma1997PRLDropPush}%
  \BibitemOpen
  \bibfield  {author} {\bibinfo {author} {\bibfnamefont {G.}~\bibnamefont {Tripathy}}\ and\ \bibinfo {author} {\bibfnamefont {M.}~\bibnamefont {Barma}},\ }\href {https://doi.org/10.1103/PhysRevLett.78.3039} {\bibfield  {journal} {\bibinfo  {journal} {Phys. Rev. Lett.}\ }\textbf {\bibinfo {volume} {78}},\ \bibinfo {pages} {3039} (\bibinfo {year} {1997})}\BibitemShut {NoStop}%
\bibitem [{\citenamefont {Tripathy}\ and\ \citenamefont {Barma}(1998)}]{TripathyBarma1998PREDropPush}%
  \BibitemOpen
  \bibfield  {author} {\bibinfo {author} {\bibfnamefont {G.}~\bibnamefont {Tripathy}}\ and\ \bibinfo {author} {\bibfnamefont {M.}~\bibnamefont {Barma}},\ }\href {https://doi.org/10.1103/PhysRevE.58.1911} {\bibfield  {journal} {\bibinfo  {journal} {Phys. Rev. E}\ }\textbf {\bibinfo {volume} {58}},\ \bibinfo {pages} {1911} (\bibinfo {year} {1998})}\BibitemShut {NoStop}%
\bibitem [{\citenamefont {Hinrichsen}\ \emph {et~al.}(1997)\citenamefont {Hinrichsen}, \citenamefont {Livi}, \citenamefont {Mukamel},\ and\ \citenamefont {Politi}}]{Hinrichsen1997Wetting}%
  \BibitemOpen
  \bibfield  {author} {\bibinfo {author} {\bibfnamefont {H.}~\bibnamefont {Hinrichsen}}, \bibinfo {author} {\bibfnamefont {R.}~\bibnamefont {Livi}}, \bibinfo {author} {\bibfnamefont {D.}~\bibnamefont {Mukamel}},\ and\ \bibinfo {author} {\bibfnamefont {A.}~\bibnamefont {Politi}},\ }\href {https://doi.org/10.1103/PhysRevLett.79.2710} {\bibfield  {journal} {\bibinfo  {journal} {Phys. Rev. Lett.}\ }\textbf {\bibinfo {volume} {79}},\ \bibinfo {pages} {2710} (\bibinfo {year} {1997})}\BibitemShut {NoStop}%
\bibitem [{\citenamefont {Hinrichsen}\ \emph {et~al.}(2000)\citenamefont {Hinrichsen}, \citenamefont {Livi}, \citenamefont {Mukamel},\ and\ \citenamefont {Politi}}]{Hinrichsen2000Wetting}%
  \BibitemOpen
  \bibfield  {author} {\bibinfo {author} {\bibfnamefont {H.}~\bibnamefont {Hinrichsen}}, \bibinfo {author} {\bibfnamefont {R.}~\bibnamefont {Livi}}, \bibinfo {author} {\bibfnamefont {D.}~\bibnamefont {Mukamel}},\ and\ \bibinfo {author} {\bibfnamefont {A.}~\bibnamefont {Politi}},\ }\href {https://doi.org/10.1103/PhysRevE.61.R1032} {\bibfield  {journal} {\bibinfo  {journal} {Phys. Rev. E}\ }\textbf {\bibinfo {volume} {61}},\ \bibinfo {pages} {R1032} (\bibinfo {year} {2000})}\BibitemShut {NoStop}%
\bibitem [{\citenamefont {Majumdar}(2010)}]{SatyaMajumdar2010Condensation}%
  \BibitemOpen
  \bibfield  {author} {\bibinfo {author} {\bibfnamefont {S.~N.}\ \bibnamefont {Majumdar}},\ }\href {https://global.oup.com/academic/product/exact-methods-in-low-dimensional-statistical-physics-and-quantum-computing-9780199574612} {\emph {\bibinfo {title} {Real-space Condensation in Stochastic Mass Transport Models}}}\ (\bibinfo  {publisher} {Oxford University Press},\ \bibinfo {year} {2010})\ p.\ \bibinfo {pages} {407}\BibitemShut {NoStop}%
\bibitem [{\citenamefont {Evans}\ \emph {et~al.}(1998)\citenamefont {Evans}, \citenamefont {Kafri}, \citenamefont {Koduvely},\ and\ \citenamefont {Mukamel}}]{EvansMukamel1998ABCmodel}%
  \BibitemOpen
  \bibfield  {author} {\bibinfo {author} {\bibfnamefont {M.~R.}\ \bibnamefont {Evans}}, \bibinfo {author} {\bibfnamefont {Y.}~\bibnamefont {Kafri}}, \bibinfo {author} {\bibfnamefont {H.~M.}\ \bibnamefont {Koduvely}},\ and\ \bibinfo {author} {\bibfnamefont {D.}~\bibnamefont {Mukamel}},\ }\href {https://doi.org/10.1103/PhysRevE.58.2764} {\bibfield  {journal} {\bibinfo  {journal} {Phys. Rev. E}\ }\textbf {\bibinfo {volume} {58}},\ \bibinfo {pages} {2764} (\bibinfo {year} {1998})}\BibitemShut {NoStop}%
\bibitem [{\citenamefont {Rajewsky}\ \emph {et~al.}(1998)\citenamefont {Rajewsky}, \citenamefont {Santen}, \citenamefont {Schadschneider},\ and\ \citenamefont {Schreckenberg}}]{RajewskySchadschneider1998RandomSequential}%
  \BibitemOpen
  \bibfield  {author} {\bibinfo {author} {\bibfnamefont {N.}~\bibnamefont {Rajewsky}}, \bibinfo {author} {\bibfnamefont {L.}~\bibnamefont {Santen}}, \bibinfo {author} {\bibfnamefont {A.}~\bibnamefont {Schadschneider}},\ and\ \bibinfo {author} {\bibfnamefont {M.}~\bibnamefont {Schreckenberg}},\ }\href {https://doi.org/https://doi.org/10.1023/A:1023047703307} {\bibfield  {journal} {\bibinfo  {journal} {J. Stat. Phys.}\ }\textbf {\bibinfo {volume} {92}},\ \bibinfo {pages} {151} (\bibinfo {year} {1998})}\BibitemShut {NoStop}%
\bibitem [{\citenamefont {Feller}(1991)}]{Feller1991ProbabilityI}%
  \BibitemOpen
  \bibfield  {author} {\bibinfo {author} {\bibfnamefont {W.}~\bibnamefont {Feller}},\ }\href {https://www.wiley.com/en-us/An+Introduction+to+Probability+Theory+and+Its+Applications%2C+Volume+1%2C+3rd+Edition-p-9780471257080} {\emph {\bibinfo {title} {An introduction to probability theory and its applications, Volume 1}}},\ Vol.~\bibinfo {volume} {81}\ (\bibinfo  {publisher} {John Wiley \& Sons},\ \bibinfo {year} {1991})\BibitemShut {NoStop}%
\bibitem [{\citenamefont {Hinrichsen}(2000)}]{Hinrichsen2000AbsorbingReview}%
  \BibitemOpen
  \bibfield  {author} {\bibinfo {author} {\bibfnamefont {H.}~\bibnamefont {Hinrichsen}},\ }\href {https://doi.org/10.1080/00018730050198152} {\bibfield  {journal} {\bibinfo  {journal} {Adv. Phys.}\ }\textbf {\bibinfo {volume} {49}},\ \bibinfo {pages} {815} (\bibinfo {year} {2000})}\BibitemShut {NoStop}%
\bibitem [{\citenamefont {\'Odor}(2004)}]{Odor2004NoneqUniversalityRevModPhys}%
  \BibitemOpen
  \bibfield  {author} {\bibinfo {author} {\bibfnamefont {G.}~\bibnamefont {\'Odor}},\ }\href {https://doi.org/10.1103/RevModPhys.76.663} {\bibfield  {journal} {\bibinfo  {journal} {Rev. Mod. Phys.}\ }\textbf {\bibinfo {volume} {76}},\ \bibinfo {pages} {663} (\bibinfo {year} {2004})}\BibitemShut {NoStop}%
\bibitem [{\citenamefont {Mahapatra}\ \emph {et~al.}()\citenamefont {Mahapatra}, \citenamefont {Bandyopadhyay},\ and\ \citenamefont {Barma}}]{Samvit2024Unpublished}%
  \BibitemOpen
  \bibfield  {author} {\bibinfo {author} {\bibfnamefont {S.}~\bibnamefont {Mahapatra}}, \bibinfo {author} {\bibfnamefont {M.}~\bibnamefont {Bandyopadhyay}},\ and\ \bibinfo {author} {\bibfnamefont {M.}~\bibnamefont {Barma}},\ }\href@noop {} {\bibinfo  {journal} {{\color{violet} {U}npublished}}\ }\BibitemShut {NoStop}%
\end{thebibliography}%

\clearpage{}

\pagebreak
\begin{widetext}
\begin{center}
\textbf{\large End Matter}
\end{center}
\end{widetext}

\begin{figure}[h]
\includegraphics[scale=0.28]{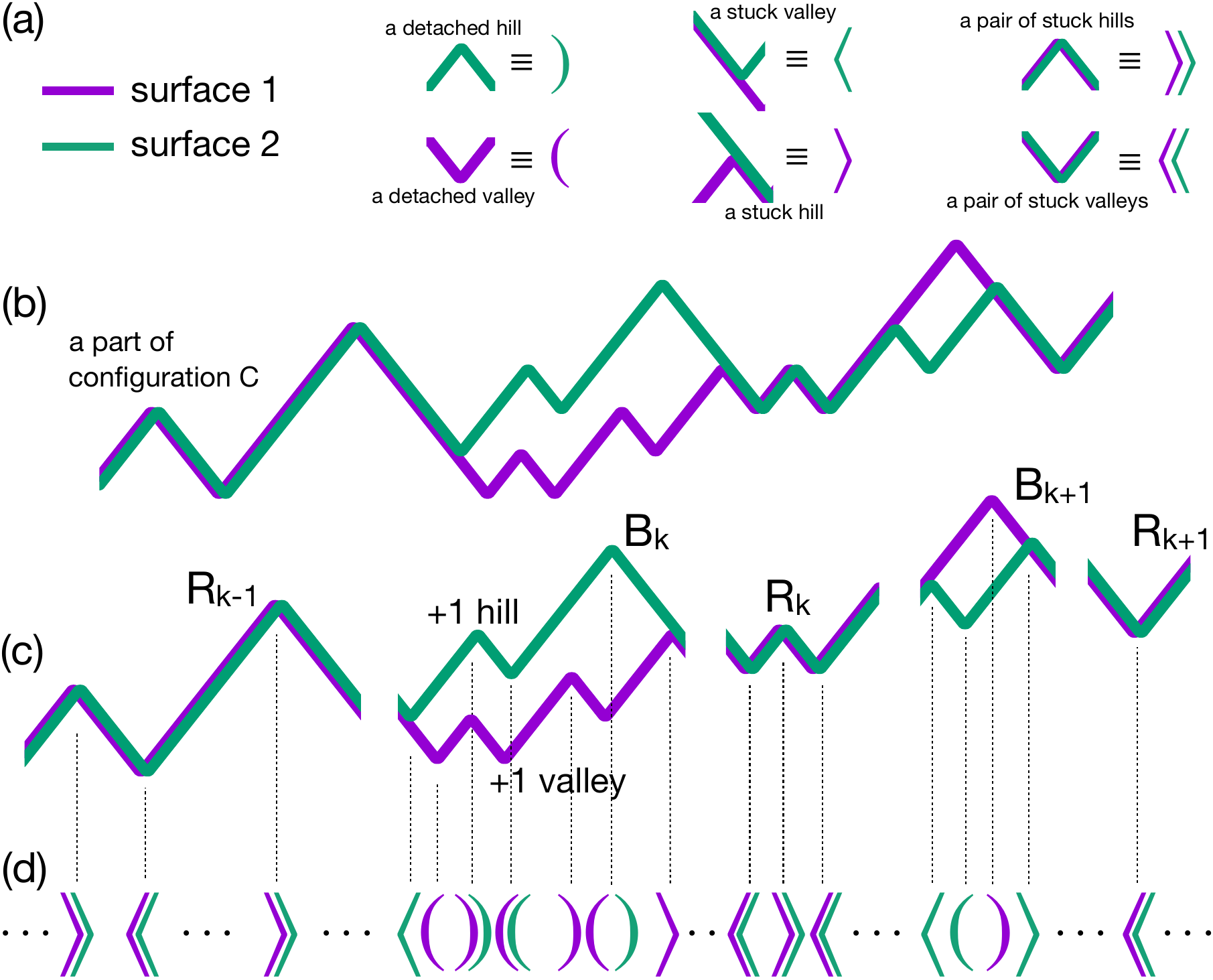}

\caption{Representing the pairing of local hills and valleys of a typical two-surface
configuration (a) using a brackets notation. (b) A part of an arbitrary
configuration $C$, whose (c) alternating bubbles $B_{i}$'s and stuck
segments $R_{i}$'s are represented separately (d) as an ordered sequence
of brackets. For clarity, the brackets arising from the two surfaces
are shown in different colors. Within each bubble, its upper surface
contains \emph{one} hill $\diagup\diagdown$ in excess of its number
of residing valleys $\diagdown\diagup$. Likewise, its lower surface
contains \emph{one} excess valley $\diagdown\diagup$. These excess
bends are paired together in our brackets notation. Since both surfaces
are identical in the 2EW and 2KPZ cases, the fluxes of these bend
pairs balance straightforwardly. }\label{fig:Brackets=000020for=000020pairwise=000020balance}
\end{figure}

EWKPZ -- Two distinct evolutions (Continued)\label{section:distinct=000020evolutions=000020cont.}---We
identify two competing processes in this regime. On one hand, the
large half-diamonds that develop from the largest bubbles $\sim O(L)$
close over timescales of $\sim O(L^{2})$ (Fig. \ref{fig:=0000202EW=0000202KPZ=000020EWKPZ=000020scaling}(e))
as the EW surface attaches onto the KPZ facet. However, the faceted
profile of the KPZ surface also facilitates a one-way detachment process
at the edges of the half-diamonds, causing the surfaces to detach
completely within ballistic timescales of $\sim O(L)/s$ \citep{SupplementalMaterial}.
The first process can compete with the second process only when $s$
is sufficiently small, heuristically when we have $s^{*}\sim O(1/L)$.
The results from simulations reflect the ballistic detachment timescales
$\sim O(L)$, however the detachment probability $s^{*}$ where the
two evolutions occur in equal fractions varies with system size $L$
as $s^{*}\sim L^{-\eta}$ with $\eta\simeq0.75$ \citep{SupplementalMaterial}.
In the 2OKPZ case, the two modes coexist only for small-sized systems.
Although diamonds also facilitate detachment, however, even relatively
small diamonds in large systems have long lifetimes, and survive without
closing to detach the surfaces completely.

Pairing scheme for the Fluxes in the 2EW and 2KPZ steady state\label{section:pairing=000020scheme=000020fluxes}---In
any configuration $C$ of the two surfaces, local updates occur only
at their bends, i.e. the local hills $\diagup\diagdown$ and valleys
$\diagdown\diagup$. The rest of configuration $C$ does not participate
in the dynamics. To (From) every bend of configuration $C$, there
exists an incoming (outgoing) flux $j_{C''\rightarrow C}$ ($j_{C\rightarrow C'}$)
from (to) another configuration $C''$ ($C'$). The configurations
$C'$ and $C''$ are nearly identical to configuration $C$, differing
only by one tilt exchange $\diagup\diagdown\rightleftharpoons\diagdown\diagup$. 

We adapt a ``brackets'' notation used in \citep{Mahapatra2020lightHeavy}
to represent configuration $C$ minimally, where only the bends of
the surfaces are represented. The round brackets $($ $)$ denote
detached bends of either surface that update with probabilities $\frac{1}{2}\pm a$,
and angular brackets $\langle$ $\rangle$ denote their stuck bends
that update with probabilities $s(\frac{1}{2}\pm a)$ (Fig. \ref{fig:Brackets=000020for=000020pairwise=000020balance}(a)).
An opening bracket $($ or $\langle$ denotes a valley, and a closing
bracket $)$ or $\rangle$ denotes a hill. In this notation, the configuration
$C$ simplifies to an ordered sequence of brackets (Fig. \ref{fig:Brackets=000020for=000020pairwise=000020balance}(b)-\ref{fig:Brackets=000020for=000020pairwise=000020balance}(d)).
The relative position of the bends in configuration $C$ determines
the ordering of the brackets in the sequence. To remove any ambiguity,
we specify that a bracket of the first surface always precedes a bracket
of the second surface whenever their bends occur at the same position.

The brackets are then grouped into pairs $(...)$ and $\langle...\rangle$
by associating every opening bracket with its nearest complementary
closing bracket. The symbol ``$...$'' indicates that these pairs
are nonlocal, i.e. they can be some distance apart, and may have other
brackets between them. Since we have periodic boundary conditions,
the numbers of stuck (detached) hills and stuck (detached) valleys
in configuration $C$ are always equal. Thus, pairing the brackets
in this manner is exhaustive. Correspondingly, in the master equation
for configuration $C$, the incoming (outgoing) flux to (from) every
opening bracket is paired with the outgoing (incoming) flux from (to)
its associated closing bracket. This pairing scheme leads to the special
form of the master equation in Eq. (1). As in \citep{Mahapatra2020lightHeavy},
the conserved tilt-exchange dynamics of our model can be recast into
a non-conserved dynamics of the brackets. The brackets notation and
pairing scheme of fluxes should be useful also for variant models
with additional features.

\begin{figure}
\includegraphics[scale=0.29]{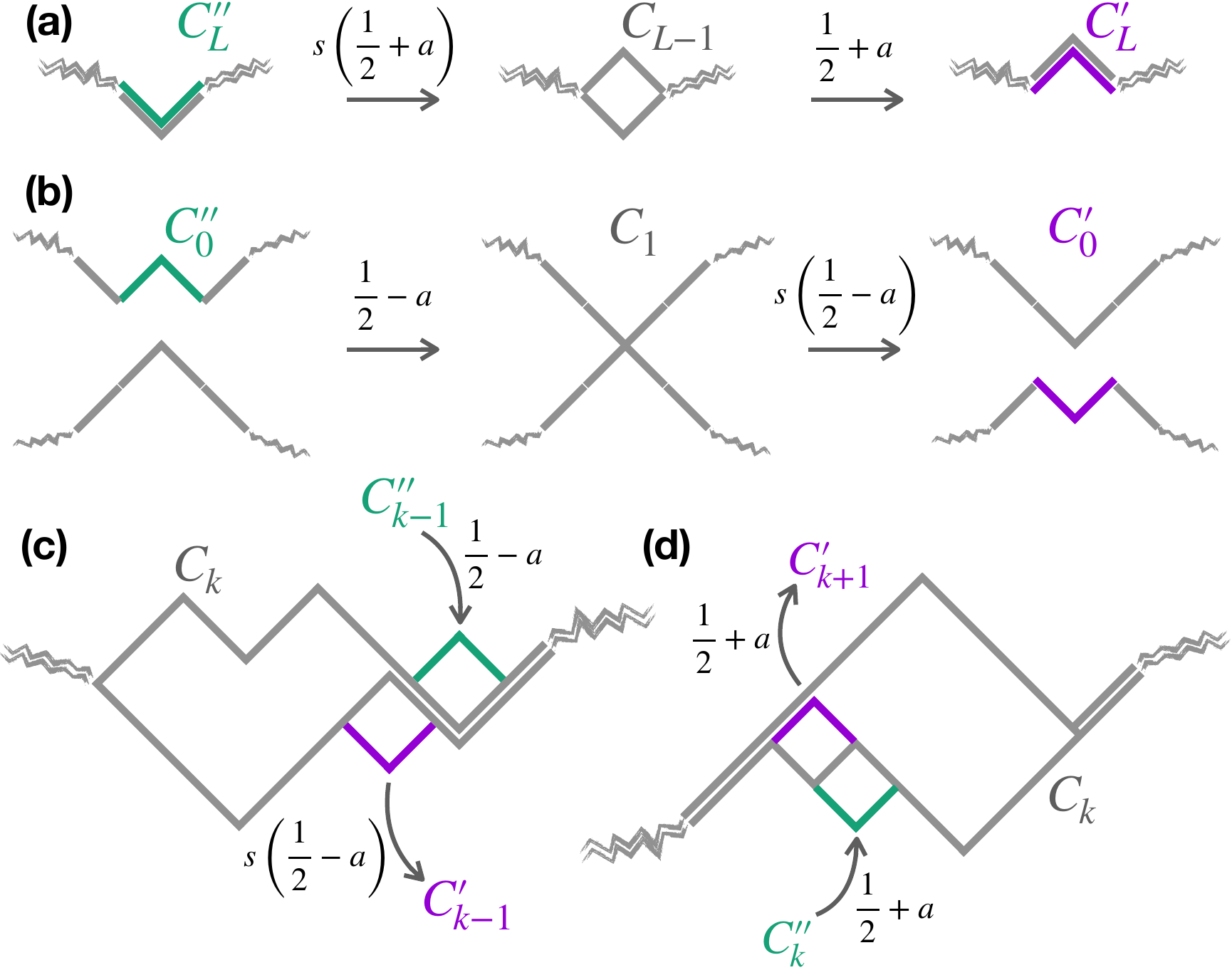}

\caption{Illustrating paired fluxes in the 2EW and 2KPZ cases for $a=a'\ge0$
in four different scenarios. The bias $a$ is oriented upwards. Around
(a) a tiny bubble in a near-completely stuck configuration of the
two surfaces $C_{L-1}^{s}$ with only one detached site. Here the
bubble closes and opens with a single microscopic update, (b) the
last stuck site in a nearly detached configuration $C_{1}^{s}$. (c)
A stuck hill at the edge of a bubble in an arbitrary configuration
$C_{k}^{s}$ with $k$ stuck sites. (d) An unstuck valley inside a
bubble of an arbitrary configuration $C_{k}^{s}$. In the four scenarios
the pairs of incoming and outgoing fluxes are (a) $(\frac{1}{2}+a)(sP(C_{L}^{''})-P(C_{L-1}))$,
(b) $(\frac{1}{2}-a)(P(C_{0}^{''})-sP(C_{1}))$, (c) $(\frac{1}{2}-a)(P(C_{k-1}^{''})-sP(C_{k}))$,
and (d) $(\frac{1}{2}+a)(P(C''_{k})-P(C_{k}))$. }\label{fig:pairwise=000020balance=000020illustration}
\end{figure}

Ultraslow timescales in the 2OKPZ case when no detachments are allowed
(s=0)\label{section:Detailed=000020balance=0000202OKPZ}---In this
case the bubbles spend a large part of their lifetimes deformed as
diamonds, keeping the surfaces adrift from any contact or pinch-off.
Therefore, the nature of contact interaction, i.e. sticking, between
the surfaces enveloping the diamonds is not significant for their
stability. Without changing our update rules, we consider a variation
of our model where the two surfaces interact through an exclusion
constraint between them instead of sticking while remaining pinned
at sites where they cross each other initially. The diamonds that
occur in this model can collapse, but their sizes remain unchanged.
It is relatively straightforward to derive their long lifetimes and
justify their stability as the update rules in this model satisfy
detailed balance exactly with respect to the long-ranged Hamiltonian
$\mathcal{H}=\epsilon\sum_{i=1}^{L}h_{1i}-h_{2i}$, defined by associating
a potential energy to the height differences between the surfaces
at every site. Then, employing activation arguments \citep{EvansMukamel1998ABCmodel,LBR2000SPS},
we can obtain the collapsing timescale of these ``non-crossing''
diamonds; this is essentially the longest rate-determining timescale
for the pinching of a diamond in our original model with sticking.

In terms of tilt variables $m_{kj}$ \citep{SupplementalMaterial},
the surface updates can be written compactly as $W(m_{1i}\leftrightarrow m_{1i+1})=\frac{1}{2}-\frac{a}{2}\left(m_{1i}-m_{1i+1}\right)$
and $W(m_{2i}\leftrightarrow m_{2i+1})=\frac{1}{2}-\frac{a'}{2}\left(m_{2i}-m_{2i+1}\right)$.
An individual update to the first surface involving a tilt exchange
$m_{1i}\leftrightarrow m_{1i+1}$ yields an energy change $\Delta E\left(m_{1i}\leftrightarrow m_{1i+1}\right)=\epsilon\left(m_{1i}-m_{1i+1}\right)$
in the Hamiltonian problem. Similarly, an update $m_{2i}\leftrightarrow m_{2i+1}$
to the second surface results in a change $\Delta E\left(m_{2i}\leftrightarrow m_{2i+1}\right)=-\epsilon\left(m_{2i}-m_{2i+1}\right)$.
If detailed balance holds, then the relation $W(C\rightarrow C_{m_{1i},m_{1i+1}})/W(C_{m_{1i},m_{1i+1}}\rightarrow C)=\mu_{ss}(C_{m_{1i},m_{1i+1}})/\mu_{ss}(C)$
must by satisfied for any two-surface configuration $C$ whose invariant
measure is given by $\mu_{ss}\left(C\right)\sim e^{-\beta\mathcal{H}\left(C\right)}$,
with configuration $C_{m_{1i},m_{1i+1}}$ differing from $C$ by the
exchange of $m_{1i}\leftrightarrow m_{1i+1}$ in the first surface.
The above relation is satisfied for $\frac{1/2-aX_{1i}}{1/2+aX_{1i}}=e^{-2\beta\epsilon X_{1i}}$
with $X_{1i}=\frac{1}{2}\left(m_{1i}-m_{1i+1}\right)$, and simplifies
to $\beta\epsilon=\frac{1}{2}\ln\left(\frac{1/2+a}{1/2-a}\right)$.
Simultaneously, the relation $\beta\epsilon=\frac{1}{2}\ln\left(\frac{1/2-a'}{1/2+a'}\right)$
arising from the dynamics of the second surface must also be satisfied.
The two relations are supported in the 2OKPZ case when we have $a=-a'$.
From the Hamiltonian $\mathcal{H}$, it is evident that the energy
of a configuration is proportional to the total area enclosed by all
its bubbles. Since diamonds have the maximum area across different
bubble profiles, therefore, configurations that possess diamonds with
$h_{1}>h_{2}$ around them are more favored for $\epsilon>0$; the
energy of such configurations is minimized and their measure $\mu_{ss}\left(C\right)\sim e^{-\beta\mathcal{H}\left(C\right)}$
maximized. Any typical configuration has large diamonds $\sim O(L)$
originating from the largest bubbles in the initial configurations;
their energy is therefore dictated by the area of the large diamonds
$\sim O(L^{2})$. On the other hand, a nearly closed configuration,
with the two surfaces in near proximity along their lengths but without
contact, has energy $\sim O(L)$. To deform into a nearly closed state,
a typical configuration with large diamonds must therefore negotiate
a super-extensive energy barrier of $\sim O(L^{2})$; this closing
process can be viewed as an Arrhenius-like activation process, leading
to long survival timescales of $\sim\exp(\lambda L^{2})$.

\clearpage{}

\widetext
\begin{center}
\textbf{\large Supplemental Materials}
\end{center}
\setcounter{equation}{0}
\setcounter{figure}{0}
\setcounter{table}{0}
\setcounter{page}{1}
\makeatletter
\renewcommand{\theequation}{S\arabic{equation}}
\renewcommand{\thefigure}{S\arabic{figure}}
\renewcommand{\bibnumfmt}[1]{[S#1]}
\renewcommand{\citenumfont}[1]{#1}

\section{Single step surfaces}\label{section:single=000020step=000020surfaces}

Each individual surface in our model is a discrete single-step surface
in 1+1 dimensions with $L$ sites, composed of an equal number of
up $\diagup$ and down $\diagdown$ tilts, therefore, $N_{\diagup}=N_{\diagdown}=L/2$
\citep{halpin1995kinetic}. The surfaces therefore have no overall
slope. In any surface configuration, the sites between every two tilts
have a certain height that is denoted by $h_{ki}$, where $k=1,2$
labels the surfaces and $i=1,2,\:...\:L$ labels their sites. The
sequence of tilts from a reference site $i=1$ determine the heights
through $h_{ki}=\sum_{j=1}^{i}m_{kj}$ with $k=1,2$. The tilt variables
$m_{kj}$ take values $\pm1$ for $\diagup$ and $\diagdown$ tilts,
respectively. On imposing periodic boundary conditions on the tilts
$m_{k1}=m_{kL}$, it follows that the boundary heights satisfy $h_{k1}=h_{kL}$,
because we have $N_{\diagup}=N_{\diagdown}$. Without interactions,
each discrete surface evolves through conserved stochastic dynamics
implemented by random sequential updates \citep{RajewskySchadschneider1998RandomSequential}.
Two neighboring tilts of the first (second) surface undergo exchanges
$\diagup\diagdown\rightarrow\diagdown\diagup$ with probability $\frac{1}{2}+a$
($\frac{1}{2}+a'$) and exchanges $\diagdown\diagup\rightarrow\diagup\diagdown$
with probability $\frac{1}{2}-a$ ($\frac{1}{2}-a'$) (Fig. \ref{fig:phase=000020diagram}(a)),
where the parameter $a$ ($a'$) is the bias of their update probabilities.
Consequently, the heights update as $h_{ki}\rightarrow h_{ki}\pm2$
and satisfy the single-step constraint $|h_{ki}-h_{ki+1}|=1$. 

Representing the tilts as particles and holes, each discrete surface
maps exactly to the simple symmetric exclusion process (SEP) for $a=0$
and to the asymmetric simple exclusion process (ASEP) for $a\ne0$.
In the continuum limit, an individual surface is described by the
Edwards-Wilkinson (EW) and Kardar-Parisi-Zhang (KPZ) universality
classes for $a=0$ and $a\ne0$, respectively. Individual EW and KPZ
surfaces, while strongly differing in their dynamics, share the same
steady state. Likewise, the steady state of a discrete single-step
surface with periodic boundaries, an equiprobable measure over all
its configurations, is independent of the bias $a$ in its update
probabilities. This $a$-invariant steady state is determined easily
by showing that the update probabilities satisfy pairwise balance
\citep{Schutz1996Pairwise,TripathyBarma1997PRLDropPush,Mahapatra2020lightHeavy}.

\section{Bubbles in the initial state and their configurational weights}\label{section:bubbles=000020in=000020initial=000020state=000020and=000020weights}

The initial state is drawn from an ensemble of random configurations
of the two surfaces. The two surfaces are placed with equal heights
at the first site, i.e. $h_{11}=h_{21}$, ensuring some overlap in
the beginning. In the relative coordinate $h_{1}-h_{2}$, every configuration
of a bubble is akin to a return trajectory of a lazy random walk that
moves rightwards (leftwards) with probability $\frac{1}{4}$ ($\frac{1}{4}$),
or stays put with probability $\frac{1}{2}$ in any time step. Thus,
the size of a bubble corresponds to the times between successive returns
of the walk to the origin. 

In a random two-surface configuration, the tilts $m_{1i}$, $m_{2i}$
at any site $i$ are either up-up, down-down, up-down, or down-up,
each occurring with probability $\frac{1}{4}$. The change in the
height difference is $0$, $0$, $+2$, and $-2$, respectively. Within
every bubble configuration, the two surfaces have different tilts
at the edge sites. For a bubble of size $l=n+2$, where $+2$ accounts
for the bubble's edge sites, its exact number of configurations is
given by the sum $B(n)=2\sum_{k=0}^{n/2}\frac{1}{k+1}^{2k}C_{k}.^{n}C_{2k}2^{n-2k}$,
which has a simpler form $4^{n+1}(n+\frac{1}{2})!/\sqrt{\pi}(n+2)!$.
The generating function of the configurations is $1-2z-\sqrt{1-4z}/2z^{2}$,
and its leading behavior for large $n$ is $\sim4^{n+2}/2\sqrt{\pi}n^{3/2}$. 

Thus for large system size $L$, bubbles of size $l$ are initially
distributed as $P(l)\sim l^{-3/2}$ with a cutoff at $L$. The typical
number of bubbles in the initial configurations scales as $\sim\sqrt{L}$
implying that a bubble in the initial state is typically of size $\sim\sqrt{L}$.
However, the largest bubble in a configuration is of the order of
the system size $\sim L$. This property is widely known as long leads
in Gambler's Ruin and follows from the statistics of the maximal term
in a sequence of random variables drawn from a Lévy distribution \citep{Feller1991ProbabilityI}. 

\begin{figure}[h]
\includegraphics[scale=0.5]{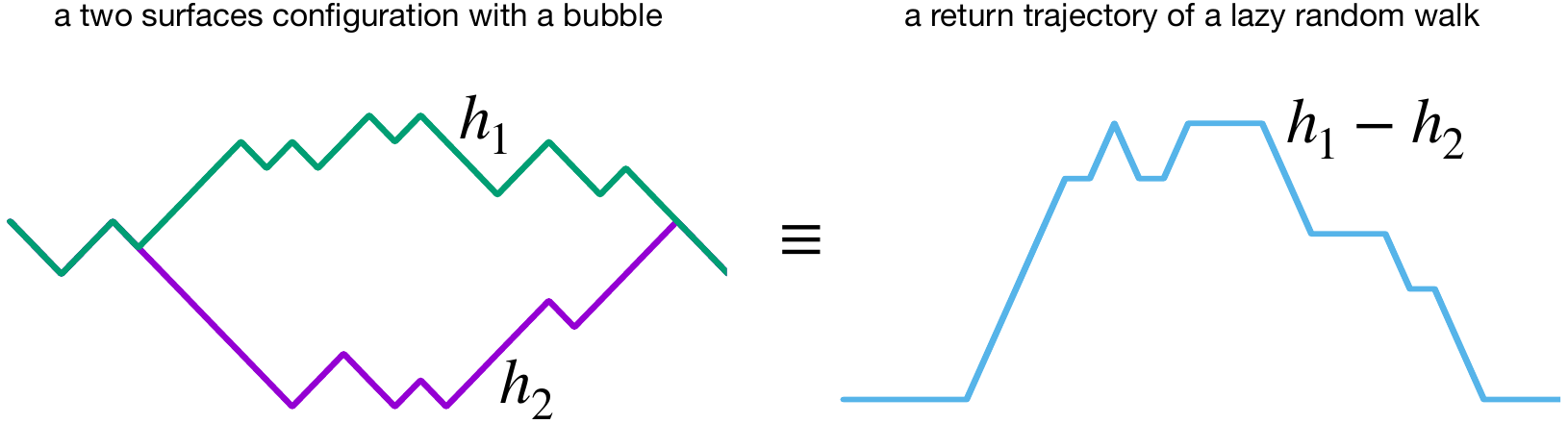}

\caption{Correspondence between bubbles and first returns of a lazy random
walk. }\label{fig:bubble=000020lazy=000020walk=000020correspondence}
\end{figure}

\clearpage{}

\section{Co-occurrence of the two evolutions in the EWKPZ case }\label{section:stability=000020two=000020evolutions}

\begin{figure}[H]
\centering
\includegraphics[scale=0.45]{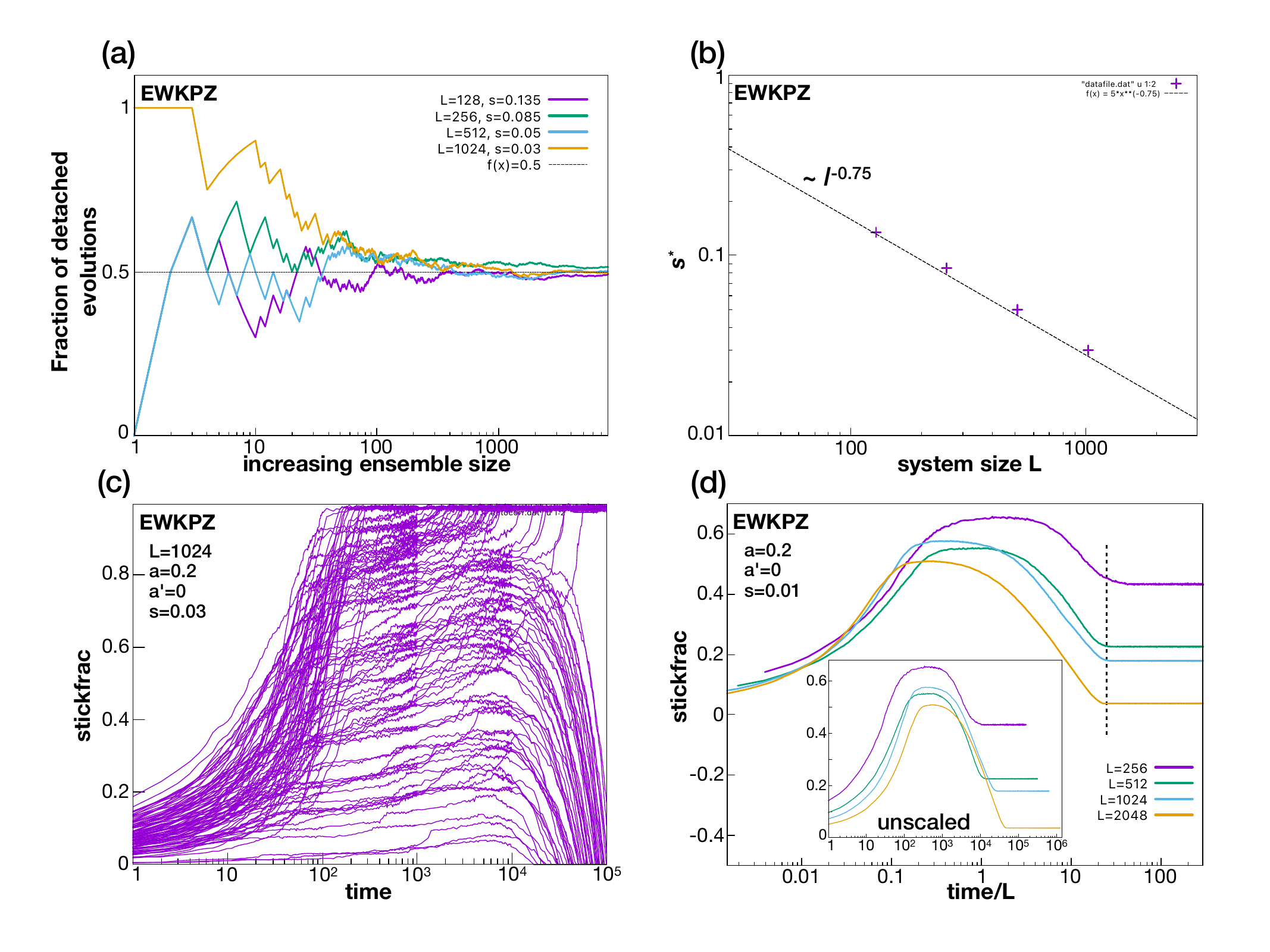}

\caption{(a) Stability of the co-occurrence of the two evolutions in the EWKPZ
case for $a=0.2$, $a'=0$. For different system sizes $L$ (=$128$,
$256$, $512$, and $1024$) and detachment probability $s^{*}(L)$
where the two evolutions occur in nearly equal fractions, the fraction
of detached trajectories versus increasing ensemble size $N_{\mathrm{ens}}$
is shown. The plot for size $L=1024$ and $s=0.03$ has the same parameter
values used in Fig. \ref{fig:ultraslow=000020dynamics=000020and=000020coexistence}(c).
(b) The detachment probability $s(L)$ vs $L$ used in (a) follows
a power law $\sim L{}^{\eta}$ with $\eta\simeq0.75$. (c) The same
plot as Fig. \ref{fig:ultraslow=000020dynamics=000020and=000020coexistence}(c)
is shown with many more sticking fraction vs time trajectories. (d)
In the EWKPZ case for non-zero $s$ ($=0.1$), the saturation of the
ensemble averaged sticking fraction vs time plots for different system
sizes $L$ aligns satisfactorily when time is scaled by $L$, in agreement
with the detachments being ballistic. The width in this case does
not collapse vertically because the fraction of the two evolutions
differs across the system sizes. For system sizes $L=$ 512, 1024
and 2048, the saturation values of the sticking fraction are small
($\lesssim0.2$); the detachment probability $s=0.1$ is relatively
large so that most evolutions culminate in detachment.}
\end{figure}

\clearpage{}

\section{Facilitated Detachment by diamonds and half diamonds }\label{section:facilitated=000020detachment}

\begin{figure}[H]
\centering
\includegraphics[scale=0.4]{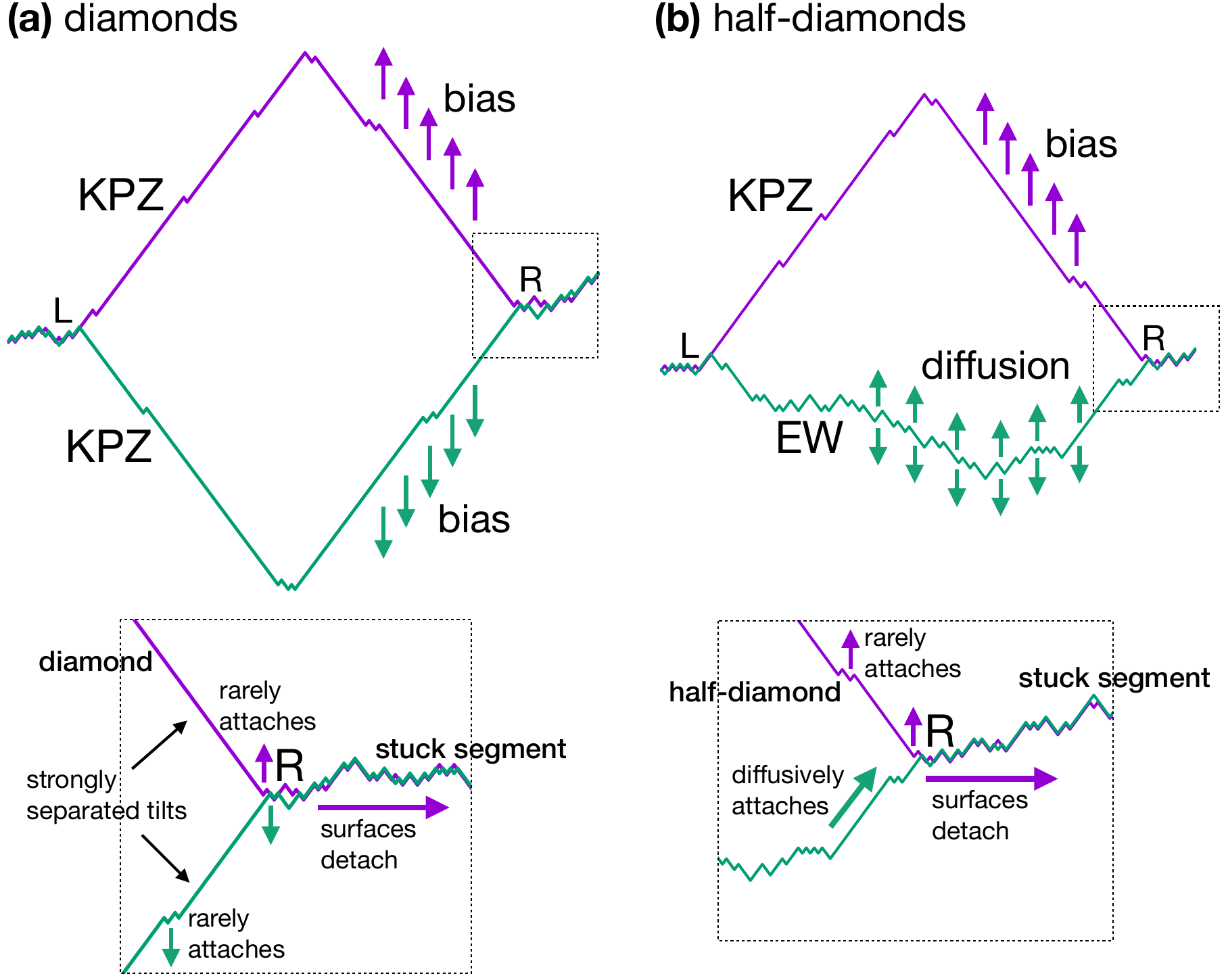}

\caption{The local hills and valleys that detach at the edges (labelled L and
R) of the KPZ facets in (a) diamonds and (b) half-diamonds are rarely
able to reattach. }
\end{figure}

The diamonds and half-diamonds, owing to their shape, facilitate the
detachment of the surfaces at their edges. Let us consider the 2OKPZ
case; for non-zero $s$ detachments occur continuously at the edges
of a diamond with a rate proportional to $s$. Whenever a stuck site
detaches at the edges of a diamond, and a local hill (valley) juts
out against the slope of its facets, it is transported quickly to
the top (bottom) of the hill (valley) due to the bias of the surfaces.
A local hill (valley) from the top (bottom) of a diamond, however,
can rarely traverse the faceted profile of a large diamond of size
$\sim O(L)$ against the bias and reach near the edges. As a result
attachments rarely occur. The predominantly one-way detachment of
the surfaces is ballistic, detaching the surfaces completely in timescales
of $\sim O(L)/s$; the detachment probability $s$ controls the speed
of this detachment process. In the EWKPZ case, the faceted KPZ surface
in the half-diamonds also facilitates detachment in a similar manner.
But when $s$ is sufficiently small in this case, the attachment of
the EW surface onto the KPZ facet competes with the detachment process.

\clearpage{}

\section{Saturation Width in 2EW and 2KPZ cases for finite sized systems}\label{section:saturation=000020width=000020finite=000020size}

\begin{figure}[H]
\centering
\includegraphics[scale=0.5]{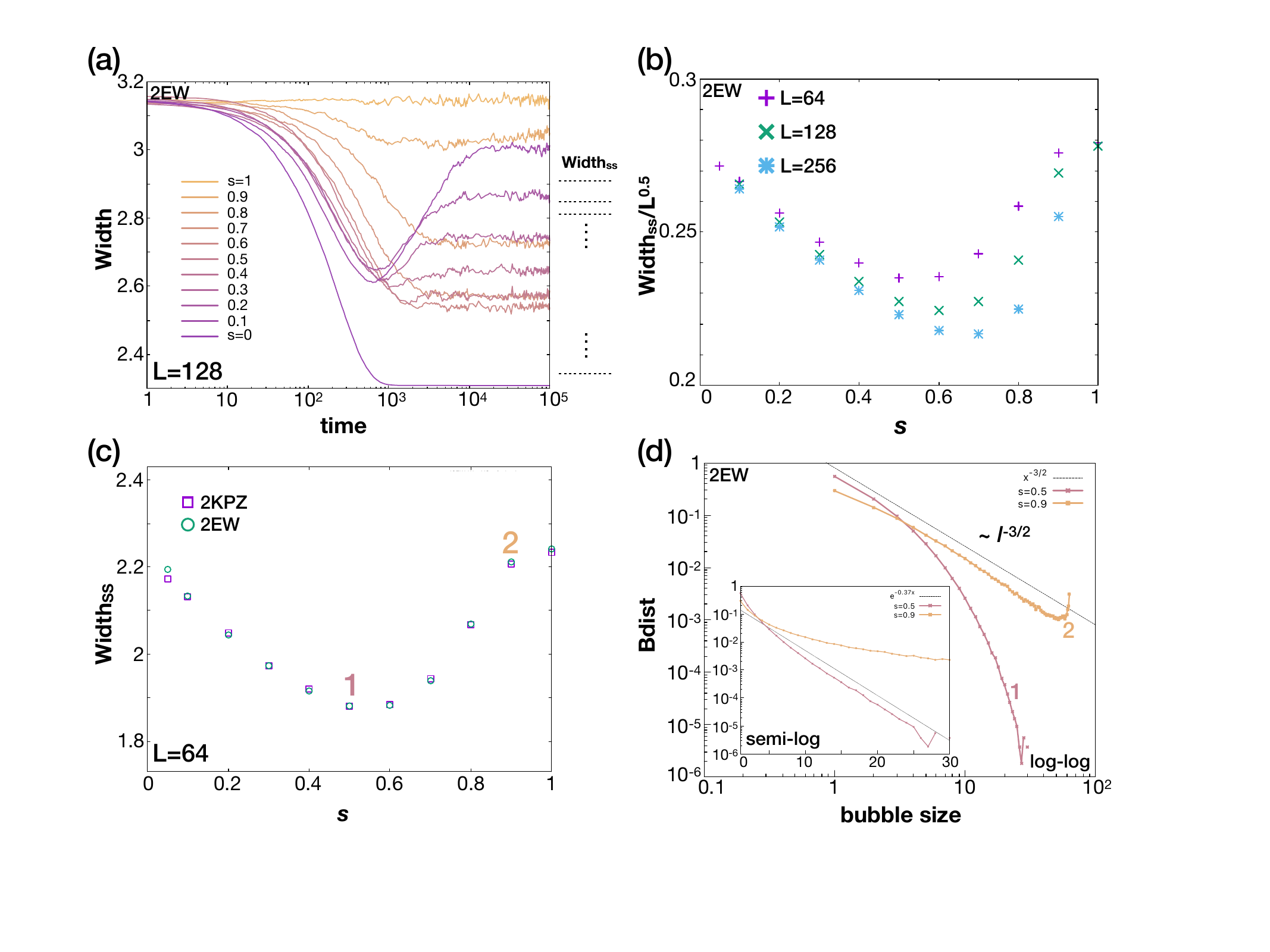}

\caption{In the 2EW and 2KPZ cases, for a finite sized system, the saturation
width of either surface shows a non-monotonicity, a minimum, with
varying detachment probability $s$. This is visible in (a) the width
vs time plots, shown in the 2EW case and system size $L=128$ for
different values of the detachment probability $s$. (b) With increasing
system size, shown for sizes $L=64$, $128$, and $256$, the detachment
probability where this minimum occurs shifts towards $s=1$. (c) In
both 2EW and 2KPZ cases shown for system size $L=64$, the saturation
width in steady state versus the detachment probability $s$ share
the same non-monotonic curve. (d) The bubble size distribution in
the steady state is qualitatively different at the minimum ($s=0.5$)
compared to when probability $s$ is close to $s=1$ ($s=0.9$); the
distribution crosses over from an exponential decay (Inset) to a distribution
that is nearly identical to a decreasing power-law with exponent $3/2$,
which holds for non-interacting surfaces. }\label{fig:saturation=000020width=000020nonmonotonicity}
\end{figure}

In finite sized systems, the plots of the saturation width in the
steady state versus the detachment probability $s$ show a non-monotonicity
shown in Fig. \ref{fig:saturation=000020width=000020nonmonotonicity},
We explain this in simple terms.

For small non-zero $s$, the two surfaces get tightly entangled and
evolve to the steady state as a single random surface. The steady
state in this limit therefore resembles that of a single surface,
i.e. a completely disordered state with equiprobable measure over
all configurations. As $s$ increases, the surfaces detach more frequently,
but now they do not fluctuate as much in concert. The surfaces do
not evolve together; instead, they get pinned to each other at different
sites. Pinning suppresses the configurations that are relatively undulating,
and as a result, the saturation width in the steady state is lowered.
When the detachment probability $s$ approaches $s=1$, the sticking
interactions become weak, and the two surfaces evolve freely; now
again they can access all configurations in the steady state with
equal probability.

\clearpage{}

\section{Dynamic exponent $z\simeq1.5$ in the 2EW case for $s=0$}\label{section:2EW=000020z=000020simeq=0000201.5=000020s=00003D0}

\begin{figure}[H]
\centering
\includegraphics[scale=0.45]{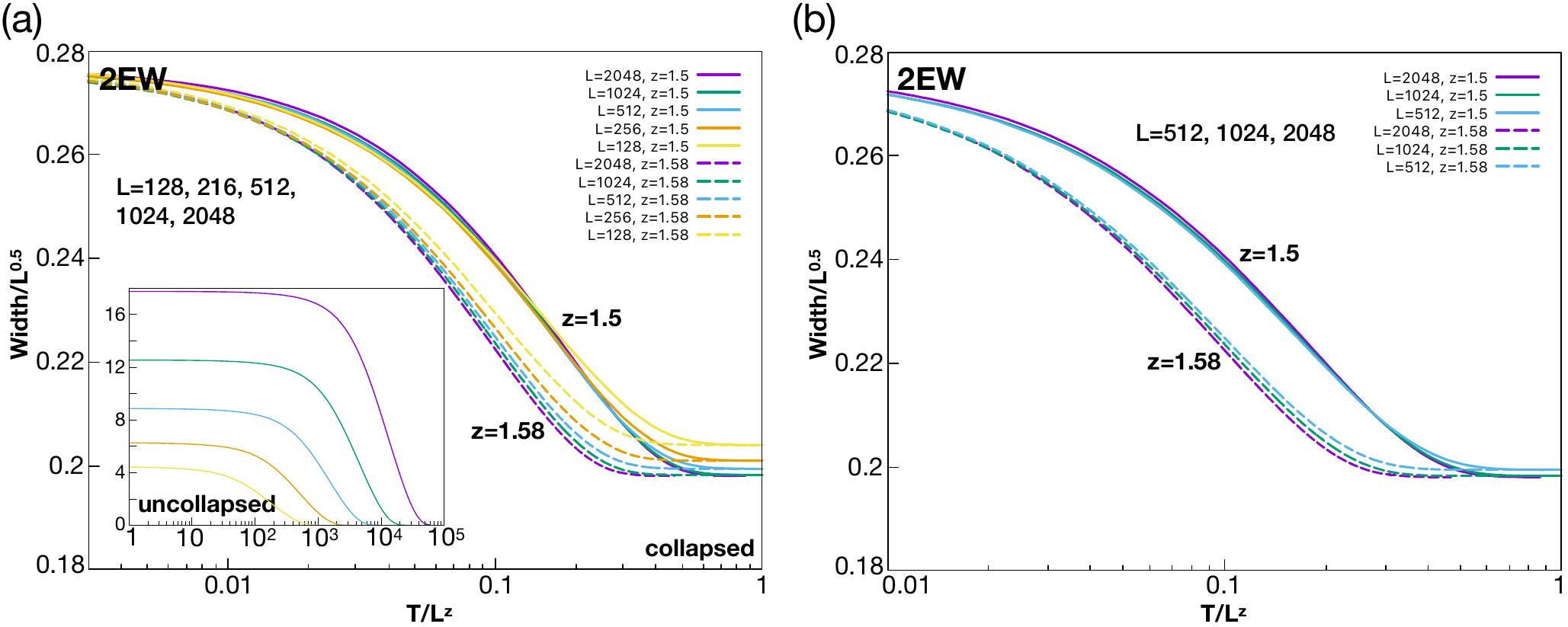}

\caption{In the 2EW case when no detachments are allowed ($s=0$), the collapse
of width vs time plots is compared when $W/L^{\alpha}$ is plotted
against $t/L^{z}$ with $\alpha\simeq0.5$ for two values of the dynamic
exponent $z\simeq1.5$ and $z\simeq1.58$. (a) The collapsed plots
are shown for system sizes $L=128$, $256$, $512$, 1024, and 2048.
In (b) the same plot as (a) is shown keeping only the three largest
system sizes $L=512$, $1024$, and $2048$. The plots are averaged
over an ensemble of size $N_{\mathrm{ens}}=50000$. }\label{fig:z=00003D1.5=000020vs=0000201.58}
\end{figure}

The origin of the dynamic (transient) exponent $z\simeq1.5$ ($\theta\simeq1.6$)
observed in the 2EW case for $s=0$ (nonzero $s$) remains to be understood.
In systems with absorbing states, for instance in models of wetting
on substrates, a dynamic exponent close to $z\simeq1.5$ is typically
attributed to directed percolation ($z\simeq1.58$ in 1d) \citep{Hinrichsen2000AbsorbingReview,Odor2004NoneqUniversalityRevModPhys}.
In our system, however, the exponent $z\simeq1.5$ originates from
the closing process of the largest bubbles $\sim O(L)$ in the initial
configurations \Citep{Samvit2024Unpublished}, which appears unrelated
to a percolation process. In the numerical plots of width vs time
shown in Fig. \ref{fig:z=00003D1.5=000020vs=0000201.58}, the collapse
of the plots appears marginally better for $z\simeq1.5$ compared
to $z\simeq1.58$, particularly for the larger system sizes $L=512$,
$1024$, and $2048$.
\end{document}